\documentclass[twocolumn]{aastex631}
\usepackage{comment}
\usepackage{ragged2e}
\received{November 19, 2023}
\revised{November 21, 2023}
\accepted{XXXX}
\submitjournal{AJ}

\shorttitle{COSMIC SETI}
\shortauthors{Tremblay, C.D. et al.}

\begin{document}

\correspondingauthor{Tremblay, C.D.}
\email{ctremblay@seti.org}

\title{COSMIC: An Ethernet-based Commensal, Multimode Digital Backend on the Karl G. Jansky Very Large Array for the Search for Extraterrestrial Intelligence}

\author[0000-0002-4409-3515]{Tremblay, C.D.}
\affiliation{These two authors contributed equally to this work}
\affiliation{SETI Institute, 339 Bernardo Ave, Suite 200, Mountain View, CA 94043, USA}
\affiliation{Berkeley SETI Research Center, University of California, Berkeley, CA 94720, USA}
\author[0000-0003-2669-0364]{Varghese, S.S.}
\affiliation{These two authors contributed equally to this work}
\affiliation{SETI Institute, 339 Bernardo Ave, Suite 200, Mountain View, CA 94043, USA}
\author[0000-0003-0216-1417]{Hickish, J.}
\affiliation{Real-Time Radio Systems Ltd., 111 Burley Road, Bransgore, Christchurch, Dorset, BH23 8AY, United Kingdom}
\author[0000-0002-6664-965X]{Demorest, P.B.}
\affiliation{National Radio Astronomy Observatory, 1003 Lopezville Rd., Socorro, NM 87801, USA}
\author[0000-0002-3616-5160]{Ng, C.}
\affiliation{SETI Institute, 339 Bernardo Ave, Suite 200, Mountain View, CA 94043, USA}
\affiliation{Dunlap Institute for Astronomy \& Astrophysics, University of Toronto, 50 St. George Street, Toronto, ON M5S 3H4, Canada}
\affiliation{Department of Astronomy, University of California, Berkeley, CA 94720, USA}
\author[0000-0003-2828-7720]{Siemion, A.P.V.}
\affiliation{SETI Institute, 339 Bernardo Ave, Suite 200, Mountain View, CA 94043, USA}
\affiliation{Berkeley SETI Research Center, University of California, Berkeley, CA 94720, USA}
\author[0000-0002-8071-6011]{Czech, D.}
\affiliation{Berkeley SETI Research Center, University of California, Berkeley, CA 94720, USA}
\author[0009-0001-8677-372X]{Donnachie, R.A.}
\affiliation{Mydon Solutions (Pty) Ltd., 102 Silver Oaks, 23 Silverlea Road, Wynberg, Cape Town, South Africa, 7800}
\affiliation{SETI Institute, 339 Bernardo Ave, Suite 200, Mountain View, CA 94043, USA}
\author[0000-0002-0161-7243]{Farah, W.}
\affiliation{SETI Institute, 339 Bernardo Ave, Suite 200, Mountain View, CA 94043, USA}
\affiliation{Berkeley SETI Research Center, University of California, Berkeley, CA 94720, USA}
\author[0000-0002-8604-106X]{Gajjar, V.}
\affiliation{SETI Institute, 339 Bernardo Ave, Suite 200, Mountain View, CA 94043, USA}
\author[0000-0002-7042-7566]{Lebofsky, M.}
\affiliation{Berkeley SETI Research Center, University of California, Berkeley, CA 94720, USA}
\author[0000-0001-6950-5072]{MacMahon, D.E.}
\affiliation{Berkeley SETI Research Center, University of California, Berkeley, CA 94720, USA}
\author[0000-0003-0804-9362]{Myburgh, T.}
\affiliation{Mydon Solutions (Pty) Ltd., 102 Silver Oaks, 23 Silverlea Road, Wynberg, Cape Town, South Africa, 7800}
\affiliation{SETI Institute, 339 Bernardo Ave, Suite 200, Mountain View, CA 94043, USA}
\author[0000-0002-9473-9652]{Ruzindana, M.}
\affiliation{Berkeley SETI Research Center, University of California, Berkeley, CA 94720, USA}
\author[0000-0002-7735-5796]{Bright, J.S.}
\affiliation{Astrophysics, Department of Physics, University of Oxford, Keble Road, Oxford OX1 3RH}
\affiliation{Breakthrough Listen, University of California, Berkeley, CA 94720, USA}
\author{Erickson, A.}
\affiliation{National Radio Astronomy Observatory, 1003 Lopezville Rd., Socorro, NM 87801, USA}
\author{Lacker, K.}
\affiliation{Berkeley SETI Research Center, University of California, Berkeley, CA 94720, USA}


\begin{abstract}
The primary goal of the search for extraterrestrial intelligence (SETI) is to gain an understanding of the prevalence of technologically advanced beings (organic or inorganic) in the Galaxy. One way to approach this is to look for technosignatures: remotely detectable indicators of technology, such as temporal or spectral electromagnetic emissions consistent with an artificial source. With the new Commensal Open-Source Multimode Interferometer Cluster (COSMIC) digital backend on the Karl G. Jansky Very Large Array (VLA), we aim to conduct a search for technosignatures that is significantly more comprehensive, more sensitive, and more efficient than previously attempted. The COSMIC system is currently operational on the VLA, recording data, and designed with the flexibility to provide user-requested modes. This paper describes the hardware system design, the current software pipeline, and plans for future development.

\end{abstract}

\section{Introduction}
The search for technosignatures---observable manifestations of technologically capable life---aims to constrain the prevalence and distribution of complex life in the Universe. The modern search for radio emissions with a spectro-temporal structure, inconsistent with the expected natural background and consistent with our understanding of electromagnetic technology, represents a probative and readily actionable search modality with current telescopes. Significant advances in real-time data analysis, driven by the reduced costs of computation, have led to the development of the Commensal Open-Source Multimode Interferometer Cluster (COSMIC) on the Karl G. Jansky Very Large Array (VLA; \citealt{evla_11}) in New Mexico, USA. The COSMIC system currently searches for narrow-band (a few Hz) drifting emissions in coherent beams aimed at individual targets of interest and an incoherent beam covering the entire primary field of view during standard proposed science programs and the observatory-led Very Large Array Sky Survey (VLASS; \citealt{VLASS}).

In the past, many dedicated facilities and programs have pioneered the search for radio technosignatures. Over its 5-year lifetime, Project Phoenix \citep{Tarter_1994,Backus_1995,Backus_2004} covered 1000--2000 stars at 1--4\,GHz with approximately 30--60 second observations per target source. The Search for Extraterrestrial Radio Emissions from Nearby Developed Intelligent Populations (SERENDIP; \citealt{Bowyer_1983,Werthimer_2001}) was one of the first initiated commensal surveys, conducted at the Hat Creek Radio Observatory at 1612\,MHz, on the Arecibo Telescope in Puerto Rico at 424--436\,MHz and, starting in 2014,
also on the Robert C. Byrd Green Bank Telescope (GBT) at 1--2\,GHz. As of 1997, the SERENDIP project had discovered over 400 anomalous signals whose origins were never conclusively determined \citep{donnelly_serendip_1998}, and the program proved the value of large-scale commensal observations.

Since the launch of the $Breakthrough$ $Initiatives$\footnote{\url{https://breakthroughinitiatives.org}}, in particular $Breakthrough$ $Listen$ in 2015 \citep{Worden:2017,Isaacson:2017}, dedicated observations with the Parkes 64\,m telescope (Murriyang) and the GBT have yielded improved sensitivities (approximately an order of magnitude over Project Phoenix) but on a similar order of 1000--2000 stars (i.e., \citealt{enriquez2017turbo,Price_2020, Gajjar_2022, Ma_2023}). These programs were also initially limited in frequency coverage; however, in the last 3 years, the search has broadened to cover the range of $\sim$800\,MHz to 12\,GHz (e.g., \citealt{Suresh_2023}), although few results have yet been published at the newly covered frequencies. Additionally, low-frequency projects have recently emerged in the Southern Hemisphere based on wide-field radio arrays; instead of searching for narrow-band drifting signals in beamformed data, these programs search for continuous signals in synthesized images' (e.g., \citealt{Tremblay_2022}). Although only a few signals of interest have yet been determined to be present in the data collected thus far (all likely to be terrestrial interference; \citealt{BLC1,Sheikh_2021,Ma_2023}), the observational campaigns on Murriyang and the GBT have set some of the most stringent limits on the search for radio technosignatures to date (i.e., \citealt{Price_2020,Sheikh_2021}).

Overall there are at least 12 radio telescopes worldwide with dedicated SETI programs, expanding the search over a variety of wavelengths, sensitivities, and approaches (e.g. \citealt{Tao_2023,Johnson_2023}. Many radio telescopes are trending toward Ethernet based digital architecture with multicasting capabilities to increase the scientific output through simultaneous commensal observation and processing with potential use by the SETI community. This includes MeerKAT \citep{MeerKAT_Dig,Slabber_2018}, the Murchison Widefield Array \citep{Morrison_2023}, and other CASPER instruments \citep{Hickish:2016} that have recently adopted the concept for commensal science and with real-time beamforming in mind. 

\begin{figure}
\centering
\includegraphics[width=0.9\linewidth]
{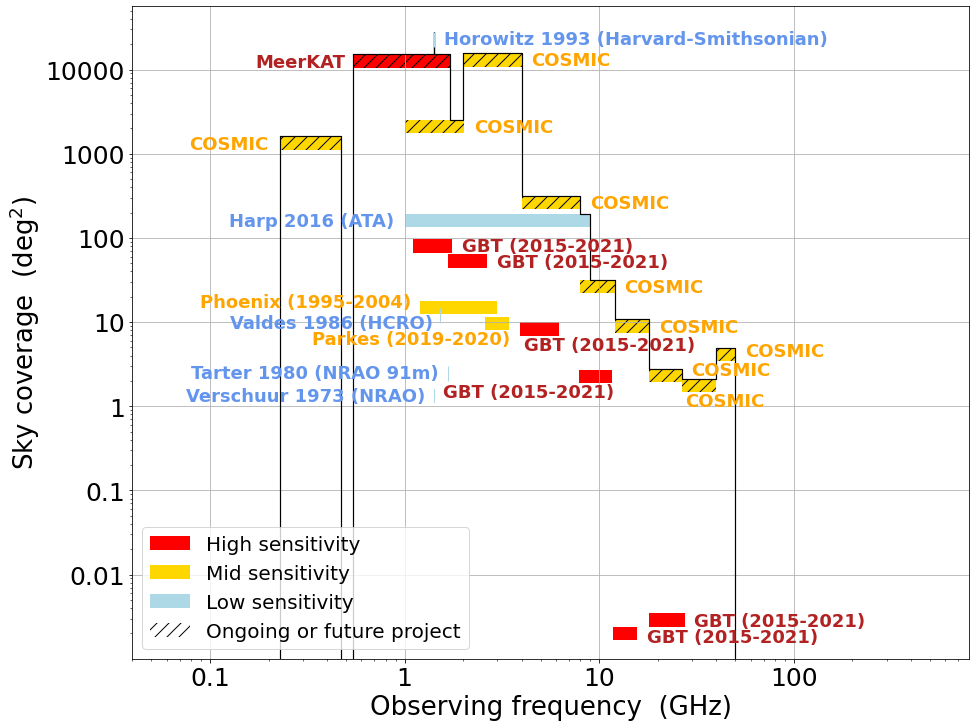}
\caption{Sky coverage vs. observation frequency for key SETI projects conducted to date (modified from Figure 5 of \citealt{Ng_2022}). The color scale represents three levels of detectability for a $10^{13}$\,W Arecibo-like transmitter emitting a 1\,Hz wide technosignature. Blue represents low sensitivity, in which the transmitter must be within 5\,pc from Earth; yellow represents medium sensitivity, corresponding to a transmitter 25\,pc away; and red represents high sensitivity, enabling detection of a source at a distance of 75\,pc.}
\label{fig:skycoverage1}
\end{figure}

Building on these foundational programs, the $SETI$ $Institute$, $Breakthrough$ $Listen$, and the $National$ $Radio$ $Astronomy$ $Observatory$ (NRAO; the scientific organization that operates the VLA) have deployed COSMIC as a new commensal equipment on the VLA bringing the multicasting digital architecture to the VLA for the first time. As a result, SETI has moved from searching a few thousand stars to searching hundreds of thousands of stars, with the potential to search tens of millions of stars over the course of a program's lifetime. Through simulations using observations from the VLA over the last three years, we can estimate the expected time on each receiver a commensal SETI program would have access to and the number of targets the corresponding search could cover \citep{Ng_2022}. As shown in Figure \ref{fig:skycoverage1}, we predict that over 10 million stars could be observed within a few years, which is orders of magnitude greater than the scope of the entire current history of radio technosignature searches \citep{Lesh_Tarter_2015}, at sensitivity levels not usually achieved in SETI experiments. These comprehensive ranges of frequencies and sky coverage are a powerful motivation for building COSMIC at the VLA.

COSMIC, as a commensal Ethernet-based backend on the VLA, receives a copy of the serial digital signals from each of the 27 operational antennas in the array, while the telescope simultaneously operates and processes the data as per standard procedures ahead of the standard VLA processing pipeline. With our copy of the data, we have the flexibility to process the data in a variety of modes, regardless of the observational frequency, region of the sky, or type of observations requested by the primary observer. This allows the system to commensally observe the sky without impacting the standard scientist-driven programs and processing. In the technosignature search mode, COSMIC executes a search targeting narrow-band (Hz-scale) emissions and produces small ``postage stamp'' raw voltage files for each antenna around signals of interest. Employing a cluster of CPU/GPU compute nodes, COSMIC processes data in real time to look for signs of technosignatures within our Galaxy from the directions of our nearest stars. However, the system is designed with significant flexibility to allow for other operational modes in the future.

\begin{figure}
\centering
\includegraphics[width=0.9\linewidth]{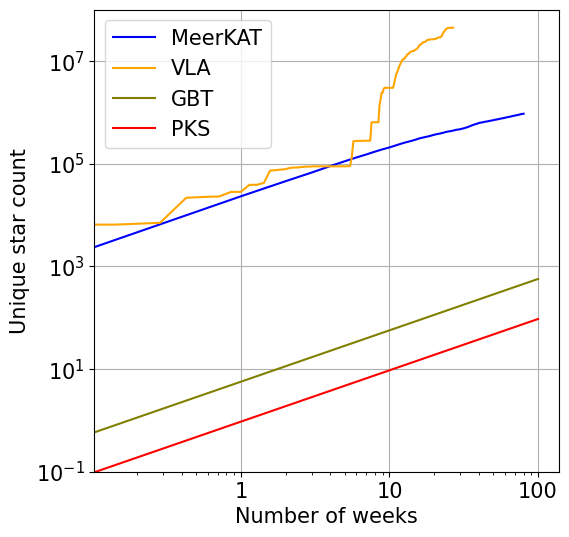}
\caption{The anticipated rate of observation for sources over the duration of the COSMIC SETI program in comparison with current programs on MeerKAT, the Murriyang telescope, and the GBT. The VLA data consist of recorded information from the VLA Low Band Ionospheric and Transient Experiment \citep{Clarke_VLITE,VLITE} and are used to simulate future observations, including commensal operation with VLASS.}
\label{fig:skycoverage}
\end{figure}

One of the major benefits of placing COSMIC on the VLA in 2023 is to conduct commensal observations along with the VLA observatory-led all-sky survey (VLASS; \citealt{VLASS}). This program, which started its third epoch in January 2023, observes the entire Northern Hemisphere at a declination (Dec) above --40 degrees and at frequencies in the range of 2--4\,GHz (S-band), enabling one of the largest sky fractions for a search for technosignatures ever attempted. By combining observations recorded during VLASS with commensal observations recorded alongside standard science-driven programs, we will cover potentially tens of millions of stars at a high sensitivity and in a frequency range of 0.074--50\,GHz (Figure \ref{fig:skycoverage}).\footnote{COSMIC does not use or calibrate the digital streams below 0.75GHz, although the VLA antennas are equipped with dipoles and digital feeds for low-frequency observations. However, future upgrades are expected to include the ability to process this frequency range on COSMIC hardware.} With these significant advances in observation scope, an answer to the question ``Are we alone in the Universe?'' may be closer than ever before.


In this paper, we provide a detailed explanation of the system hardware, software, and data processing pipeline of COSMIC. We also discuss the observation process using the on-the-fly mode at the VLA (including during VLASS), and describe future plans for COSMIC.

\section{COSMIC Hardware Overview}
In the original vision for COSMIC, as a commensal backend for the VLA, it was thought that it would run alongside the facility instrument \textsc{WIDAR}, receiving a copy of VLA antenna voltage data through a preexisting ``spigot'' connector provided by \textsc{WIDAR}'s ``baseline boards'' \citep{COSMIC, widar-carlson}. It was later determined that connecting to the VLA in this manner would have various drawbacks in terms of making COSMIC as independent as possible from the VLA's maintenance and operation. Instead, a mechanism was sought to generate a copy of the VLA data before they enter the \textsc{WIDAR} ``station board'' processors.

At the VLA, the antenna signals are digitized at each dish and transmitted via a fiber backhaul to the array's central processing facility (operations building). This unidirectional data link, known as the \emph{Digital Transmission System} (DTS; \citealt{memo33-dts-protocol}), provides an ideal location for COSMIC to source data from. By amplifying and splitting the fiber optic connection from each antenna, it became possible to generate two copies of each antenna's DTS stream, with one copy continuing to drive the \textsc{WIDAR} instrumentation and the other available for arbitrary signal processing by COSMIC.
When sourced in this way, the COSMIC input data are unaffected by all VLA primary user configurations except choices regarding analog local oscillator (LO) tuning (frequency band).
Furthermore, the data received by COSMIC are the raw ADC (analog-to-digital converted) samples from each antenna, providing maximum flexibility for processing choices in COSMIC. The resulting incoming data rates that COSMIC needs to accept are as follows:
\begin{enumerate}
\item 2\,GHz * 28 antennas * 2 polarizations * 2\,Nyquist * 8\,b = 1.7 Tb/s (for 8-bit mode);
\item 8\,GHz * 28 antennas * 2 polarizations * 2\,Nyquist * 3\,b = 2.7 Tb/s (for 3-bit mode);
\end{enumerate}

With the DTS data streams chosen as the data source for COSMIC, the system is naturally separated into four main parts:
\begin{itemize}
\item Optical Interfacing -- splitting and physically adapting the DTS fiber optic streams such that they are fed into an off-the-shelf (commercially available) high-throughput processing board.
\item Station Processors -- field-programmable gate array (FPGA) modules tasked with processing the broadband data streams for each operational antenna into narrow-band channels (multiple discrete subbands), including compensation for signal path delays and the VLA's LO tuning offset scheme.
\item Data Interconnects -- 100\,Gb\,s$^{-1}$ Ethernet switches used to implement the ``corner-turn'' operation required to rearrange the data from parallel by antenna to parallel by frequency order.
\item Array Processors -- CPU and GPU processors on which data are processed for each narrow-band channel, for each polarization, from all antennas in the VLA\footnote{Although there are 27 antennas in each VLA configuration, the NRAO has 28 antennas in total, with one being out for maintenance. We therefore have sufficient electronics to accommodate 28 independent antennas.}, where this processing includes forming phased-array beams, visibility matrices, and any other desired data products.
\end{itemize}

\begin{figure*}
    \centering
    \includegraphics[width=\textwidth, height = 7cm]{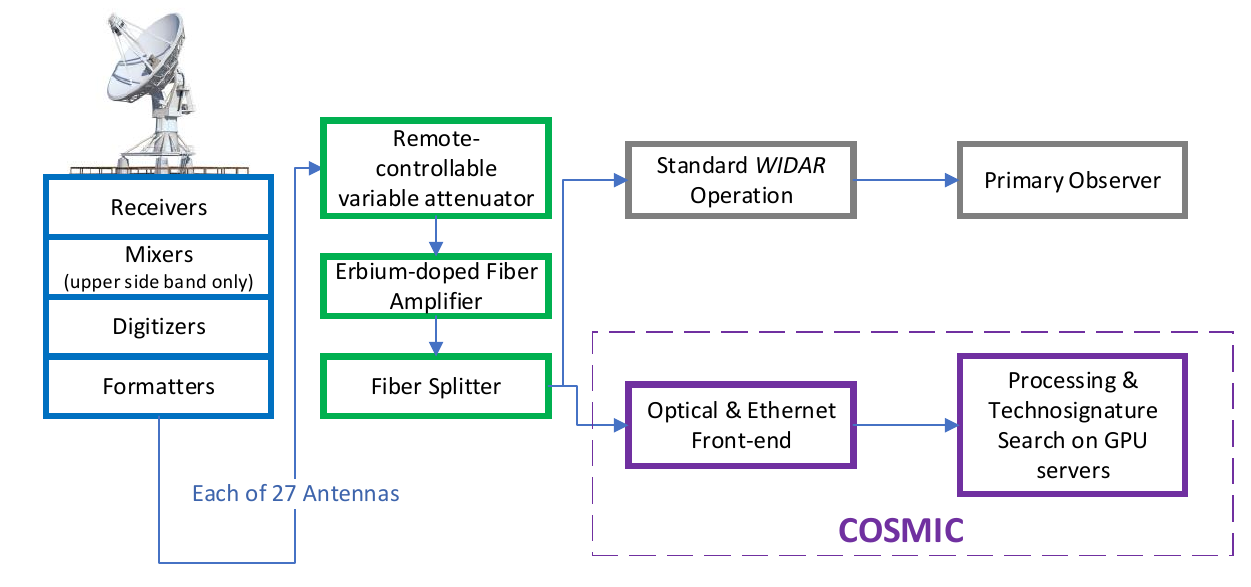}
\caption{A high-level diagram of the data flow from the VLA antennas in the context of COSMIC. The signals from each antenna pass through the receivers, and in the 8-bit mode (the VLA can operate in both a 3-bit and an 8-bit mode), all received frequencies are shifted to the X-band (8--10\,GHz; upper side band). COSMIC receives one copy of the digitized signals after they are amplified through an erbium-doped fiber amplifier (EDFA). The other copy is processed through the \textsc{WIDAR} correlator for standard VLA processing. \\}
\vspace{-0.5em}
    \label{fig:widar_split}
\end{figure*}

The top-level architecture of the COSMIC system is shown in Figure \ref{fig:widar_split}, and the individual parts of the system design are described in the remainder of this section.

\subsection{Optical Interfacing}
Each antenna in the VLA outputs data via the VLA DTS, as described in depth in \citealt{memo33-dts-protocol, memo420-dts-protocol}.
For each antenna, digitized voltage data for the two polarizations and multiple frequency tunings (i.e., intermediate frequencies, LO tunings or IFs) are framed, augmented with timing and error-check metadata, and transmitted over a single fiber.
Each DTS fiber carries 120\,Gb\,s$^{-1}$ of data---consisting of 96\,Gb\,s$^{-1}$s of digitized sample data plus protocol overload---over 12 wavelength-multiplexed 10\,Gb\,s$^{-1}$ lanes, which utilize $\sim$1550\,nm lasers spaced at 200\,GHz.
The DTS protocol is nonstandard but has the following features:
\begin{itemize}
\item Synchronization patterns are built into the protocol to allow a downstream receiver to perform clock recovery and read the data stream without additional reference clocks.
\item The underlying data transport bit clock is synchronous with the VLA samplers, enabling a receiver to recover the original sampling clock.
\item A timing pulse every 50\,ms is embedded in the DTS streams, allowing streams from multiple antennas to be deterministically aligned in time.
\item Checksums (the number of bits in a transmitted message) are transmitted within the data streams to enable a receiver to monitor transmission errors.
\end{itemize}

Since the DTS protocol has a custom design, it is readable by a low-level programmable logic chip, such as an FPGA.
The role of the COSMIC optical interfacing system is to split the DTS fibers and manipulate the COSMIC copy of the DTS stream to interface with an off-the-shelf FPGA processing platform. At each antenna in the VLA, two intermediate frequency (IF) bands exist, with a total bandwidth of up to 2\,GHz in the 8-bit mode and 8\,GHz in the 3-bit mode. Currently, COSMIC operates only when the telescope is operating in the 8-bit mode, utilizing up to 1024\,MHz of bandwidth for each IF.

\subsubsection{DTS Splitting}
Each of the VLA antenna's DTS fibers is split with inexpensive optical components.
However, it is of paramount importance that the fiber splitting (and resulting $\sim3$\,dB attenuation of the DTS signal at each of the splitter outputs) does not adversely affect the ability of \textsc{WIDAR} or COSMIC to receive the data streams.
To maintain the DTS power levels after splitting and, further, to allow flexible tuning of these levels to support the VLA configuration-dependent input power levels, the COSMIC deployment includes a tunable variable attenuator and fixed amplification of the DTS signals upstream of the fiber split.

\begin{figure}
    \centering
    \includegraphics[width=0.48\textwidth]{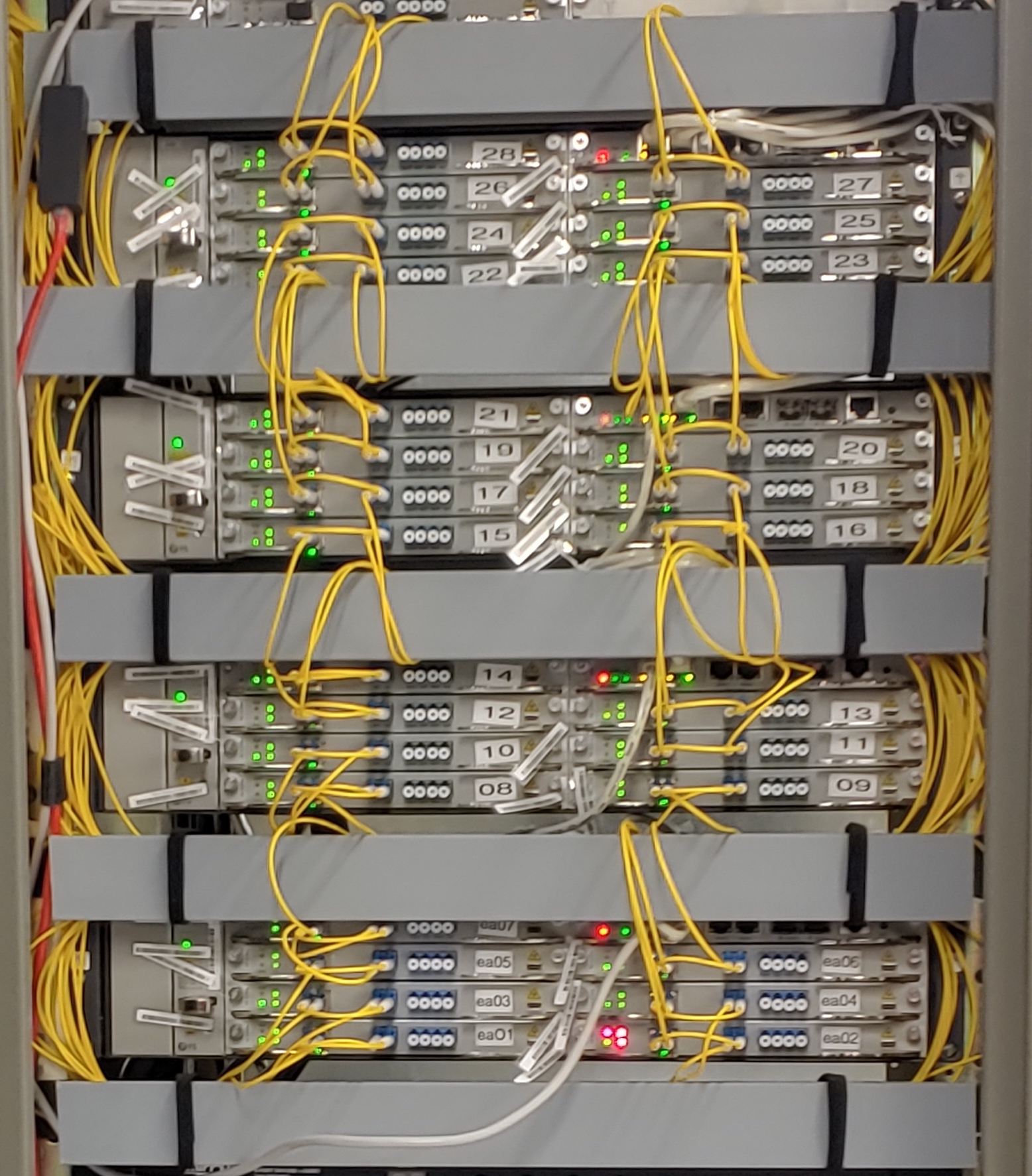}
\caption{The signal attenuators, amplifiers and splitters installed for the COSMIC system in the VLA \textsc{WIDAR} correlator room. All of them are labeled with their corresponding antenna numbers.}
\vspace{-0.5em}
    \label{fig:amplifier-splitter}
\end{figure}

After a period of testing and qualification, the Fiberstore M6200-25PA erbium-doped fiber amplifier (EDFA) was chosen for COSMIC and installed by the NRAO (Figure \ref{fig:amplifier-splitter}).
This amplifier is a cost-effective, off-the-shelf product providing 25\,dB of amplification and designed to support long-range wavelength-multiplexed data links.
The M6200-25PA\footnote{\url{https://www.fs.com/products/107367.html}} EDFA also supports the addition of a remote-controllable variable attenuator.
One amplifier is required for each VLA antenna, and these are housed in rack-mounted enclosures, with each enclosure supporting up to 7 individual amplifiers (Figure~\ref{fig:amplifier-splitter}).

\subsubsection{Electrical Conversion}
Following the DTS split, the 12 multiplexed optical carriers on each fiber must be separated and converted into an electrical form with which the downstream electronics can interface.

A standard passive wavelength demultiplexer is used, which is factory-configured to target the precise laser wavelength used at the VLA.
Once a link is separated into individual fibers, these are fed into standard Quad Small Form-factor Pluggable Plus (QSFP+) 40GBASE-PLR4 optical transceiver modules.
These modules are designed to receive the four parallel $\sim10$\,Gb\,s$^{-1}$ optical data streams that make up a 40\,Gb\,s$^{-1}$ Ethernet link.
While these modules are designed to operate with 1310\,nm optical carriers, testing has shown that their performance is not significantly degraded when they are used as receivers for the VLA's 1550\,nm signals.
As each QSFP+ transceiver is capable of converting four optical channels into electrical signals, three transceivers are required to convert all the signals from a single VLA antenna (Figure \ref{fig:cosmic-dts-rack}).

\subsection{Station Processors}
The role of the COSMIC station processors is to receive the DTS data streams, provide station-level data processing, and output data as a stream of User Datagram Protocol/Internet Protocol (UDP/IP) packets to the downstream array processing system.
One of the driving goals of the COSMIC implementation is to use as much off-the-shelf equipment as possible to minimize nonrecurring engineering costs and risks.
For the COSMIC station processors, this meant finding a cost-effective commercial platform offering the following features:
\begin{itemize}
\item A powerful FPGA capable of interfacing with the custom DTS data format and providing sufficient computing resources for digital signal processing.
\item Many QSFP+ connections to facilitate interfacing with the 336 input data fibers (12 per antenna) as well as further connections for outputting processed data.
\item A form factor based on rack-mountable enclosures to minimize the need for thermal and mechanical engineering.
\end{itemize}

\begin{figure}
        \centering
        \includegraphics[width=0.4\textwidth]{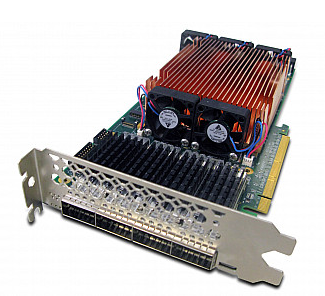}
\caption{The AlphaData ADM-PCIe-9H7 FPGA card, which supports a powerful AMD XCVU37P FPGA in a PCIe form factor. The platform provides 4 on-board QSFP+ connections, with the option to add 8 more via a pair of AD-PCIE-FQSFP quad-QSFP+ PCIe cards.}
\label{fig:FPGAcard}
\end{figure}

\begin{figure}
    \centering
    \includegraphics[width=0.45\textwidth]{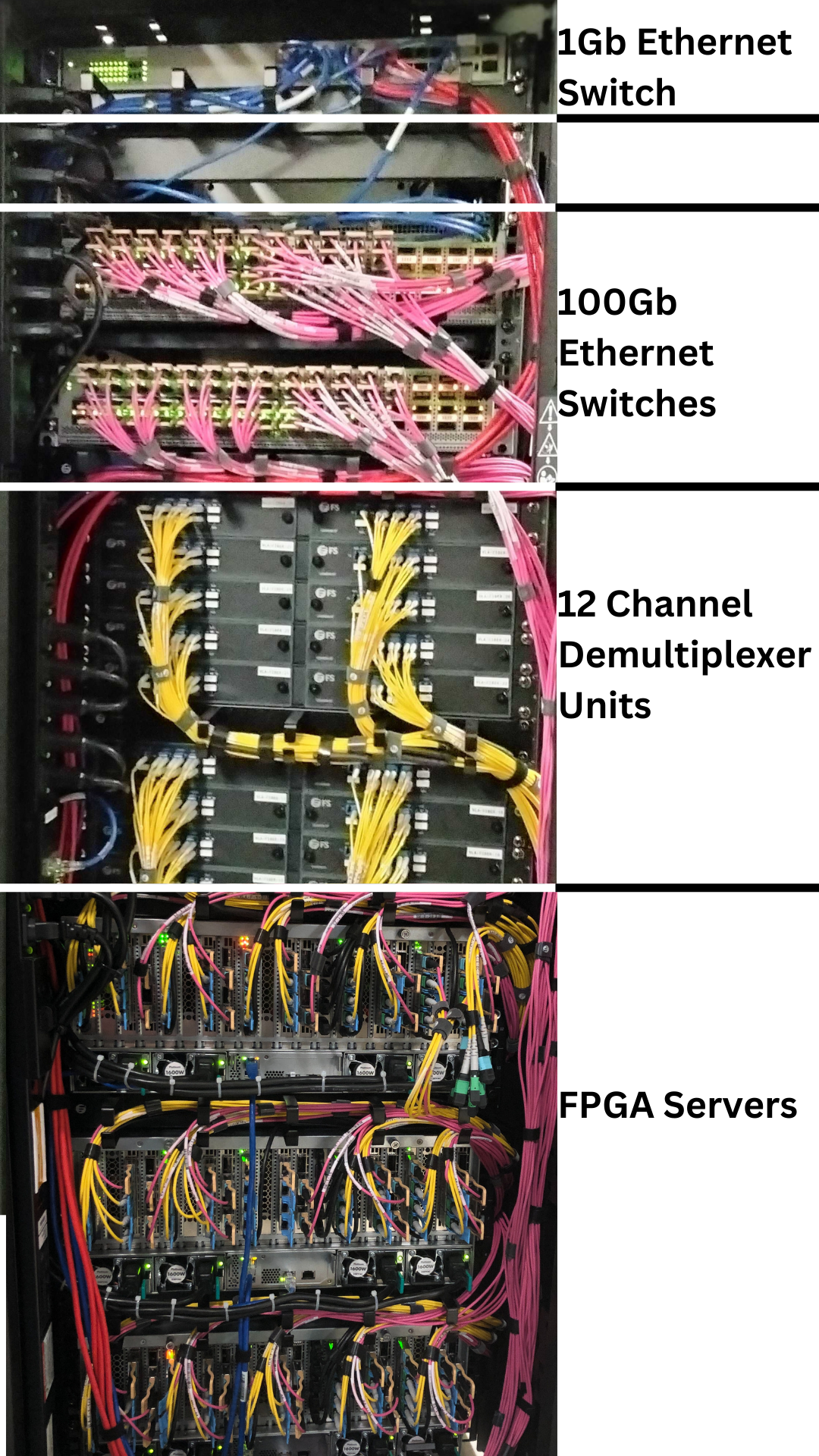}
\caption{Back view of the COSMIC DTS rack with each of the components. The image shows the 1\,Gb network switch, the primary processing node (head node), and the two 100GbE switches. The middle panel shows 10 of the 30 demultiplexer units, and the bottom panel shows the back view of the three FPGA servers hosting 15 FPGA cards, into which the 40 and 100\,Gb QSFP transceivers are plugged.
}
\vspace{-0.5em}
    \label{fig:cosmic-dts-rack}
\end{figure}

\begin{figure}
    \centering
    \includegraphics[width=0.48\textwidth]{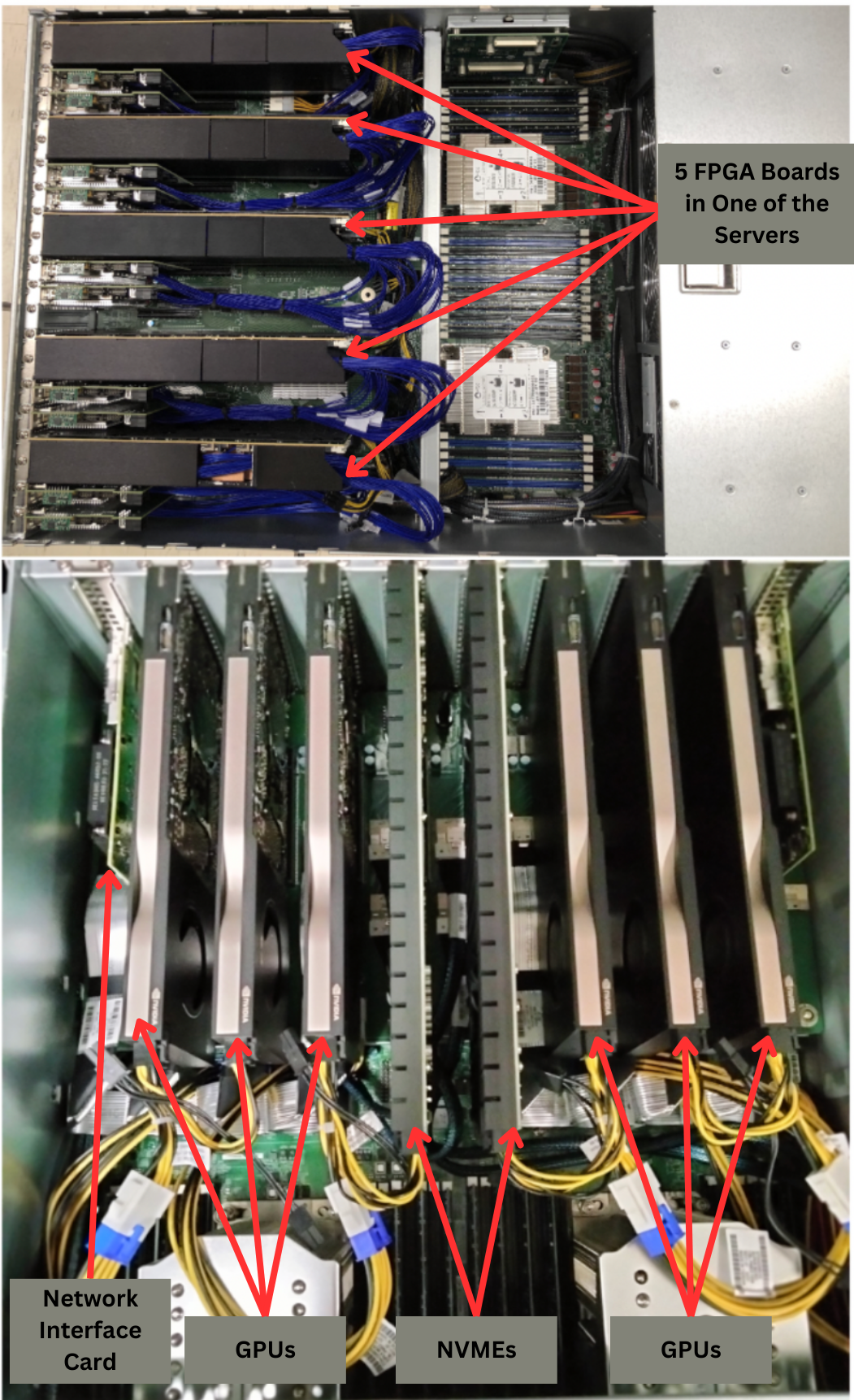}
\caption{Top: An FPGA server populated with five FPGA cards. Each FPGA board is connected to a pair of quad-port QSFP+ add-on cards, allowing it to interface with a total of 12 QSFP+ modules. This is sufficient to handle the data from two VLA antennas. Bottom: The internal components of a single GPU compute server, comprising 6 NVIDIA RTX A4000 GPUs, 2 Highpoint SSD7540 NVMe RAID cards, and 2 dual-port 100GbE network interface cards.}
\vspace{-0.5em}
    \label{fig:fpga-server}
\end{figure}

The AlphaData ADM-PCIe-9H7\footnote{\url{https://www.alpha-data.com/product/adm-pcie-9h7/}} meets all of these requirements (Figure \ref{fig:FPGAcard} and Figure \ref{fig:fpga-server}).
It is based on an AMD Virtex Ultrascale+ xcvu37p FPGA \citep{ultrascale} and provides substantial signal processing resources (including 9024 multiplier cores, 340\,Mb of on-die memory, and 8\,GB of on-chip memory) and QSFP+ connectors (4 on-board, plus eight available via AD-PCIE-FQSFP expansion cards/daughter boards).
The ADM-PCIe-9H7 comes in an industry-standard PCIe form factor and is installable in most standard rack-mountable computer servers that support GPUs.

The COSMIC system uses one ADM-PCIe-9H7 card, paired with two ADM-PCIe-9H7 quad-QSFP+ daughter cards, to process data from a pair of VLA antennas.
In this configuration, a set of cards receives 24 lanes of DTS data via six QSFP+ interfaces and transmits up to 400\,Gb\,s$^{-1}$ of data to the downstream processing system via a further four QSFP+ outputs.

The COSMIC system comprises of 15 cards to facilitate the processing of 28 antenna inputs, including the provision of two ``hot spares''.
The FPGA cards are hosted on Tyan B7119F77V10E4HR-2T55-N servers\footnote{We note that PCIe is used only for monitoring and control, and there is no high-speed interboard communication across the PCIe bus.}, which are designed to support up to 21 single-width PCIe cards in a 4U form factor.
A single server is able to support five FPGA processor cards and their QSFP+ expansions (see Figure \ref{fig:fpga-server}).

Control of the FPGA cards is exposed to the rest of the COSMIC system via a REST (a set of architectural constraints used by developers) interface running on the FPGA host servers, with communications to the FPGAs running over the PCIe bus by means of Linux drivers provided by the FPGA vendor, $Xilinx/AMD$.

\subsubsection{Station Processing}
The FPGA station processing pipeline comprises the following per-antenna actions:
\begin{itemize}
\item Receive DTS data streams and decode these streams using existing VLA firmware provided by NRAO.
\item Compensate for delays due to both signal propagation from the astronomical source of interest to the VLA antennas and fiber optic signal propagation from the antennas to the COSMIC system.
\item Remove the per-antenna VLA LO tuning offsets used by WIDAR to reject interference suffered by the VLA's IF system.
\item Divide the broadband DTS data streams into multiple 1\,MHz wide frequency bins (coarse channels).
\item Track the phases and delays of the signals in each 1\,MHz channel to compensate for sky rotation over the course of an observation.
\item Form UDP/IP packets containing a subset of frequency channels and transmit these data packets via a 100\,Gb Ethernet (100GbE) switch to a runtime-determined downstream processing node.
\end{itemize}

\subsection{Data Interconnects}
The job of the COSMIC data interconnect system is to facilitate the transfer of antenna data from the station boards to the compute cluster. This interconnect system enables the ``corner-turn'' operation frequently employed in radio astronomy, in which data streams that are arranged in parallel by antenna are converted to instead be parallel by frequency.
After this conversion, the downstream processing nodes can receive data from all antennas in the array. For COSMIC, however, only a subset of the total observation band is directed to each processing node.
These data may be correlated, beamformed, or otherwise combined to leverage the distributed nature of the VLA's observing aperture.

As is typical of many modern radio telescopes, COSMIC uses off-the-shelf Ethernet switches as data interconnects \citep{Hickish:2016}.
Such hardware allows the corner-turn operation to be accomplished by simply addressing data frames entering the network to an appropriate destination processor such that data frames containing common frequency channels end up at a common processing node.
With each COSMIC station processor outputting data over a pair of 100GbE links per antenna, the full COSMIC data interconnect system comprises 58,100GbE input streams and a potentially similar number of output streams.

For the greatest cost-effectiveness and availability of the switches, it is preferable to limit the COSMIC data network so that no more than 64 nodes in total need to connect to any one switch.
COSMIC achieves this by dividing the data over two separate 100GbE networks by ensuring that the station processors direct half of their total output bandwidth down each of their two available 100GbE outputs (one switch for each LO tuning).

With this division, the COSMIC data interconnect system can be implemented as two completely independent networks, each with 28 inputs and a similar number of outputs. Each of these networks is built around a single 64-port 100GbE switch. Such switches are widely available; specifically, COSMIC utilizes the N9K-C9364C switch, manufactured by Cisco.

\subsection{Array Processors}
The specifications of the compute cluster that processes the signals after COSMIC digital signal processing are shown in Table \ref{tab:specs-compute} and the bottom of Figure \ref{fig:fpga-server}. The high data rate of the VLA and the real-time calibration and technosignature search goals set the main requirements for the compute cluster design. The data are transmitted through a pair of 100GbE optical transceivers to each compute node in the cluster, where there are two nodes per GPU server. To facilitate real-time processing, each GPU node is allocated 32 $\times$ 1\,MHz channels out of the total 1.024\,GHz bandwidth for further channelizing, beamforming, and search processes.

In the current incarnation of COSMIC, the GPU compute cluster consists of 22 GPU servers, where each node is fitted with two network interface cards (NICs), 8\,TB of nonvolatile memory (NVMe) storage, CPUs, and GPUs. Including a small amount of overhead, each NIC on each node ingests data at a rate of $\sim$3.4\,Gb\,s$^{-1}$ during on-the-fly mapping scans, with 32\,MHz of bandwidth distributed to each of the 22 servers. In the technosignature beamforming and search mode (\S 3.7), data are transferred from the FPGAs to the NVMe storage access buffers via a 100GbE switch, as the NVMe drives provide fast read and write speeds. This design is based on benchmarks from commissioning, which indicate that a network bandwidth of at least 75\% of line-rate 100GbE can be ingested, making this a compute-limited SETI experiment.

\begin{table}
\raggedright
\caption{Configurations of the 22 compute nodes and 2 storage nodes implemented in the current design of COSMIC. See the bottom of Figure 7 for an image of the inside of one of the chassis. }
\label{tab:specs-compute}
\begin{tabular}{|p{0.75in}p{0.25in}p{1.85in}|}
\hline\hline
\textbf{Compute node} & &Details \\
\hline\hline
Chassis     &  &Supermicro 4124GS-TNR 4U GPU server\\
CPU         & 2& AMD Epyc 7313 7413 CPU \\
GPU         & 6& PNY RTX A4000 GPU \\
NIC         & 2& Mellanox MCX623106AS-CDAT dual-port 100GbE\\
Memory      & 8& 1 TB Samsung 980 Pro NVMe \\
            & & 512 GB DDR memory \\
            & 2& Highpoint SSD7540 NVMe RAID\\
\hline\hline 
\textbf{Storage node} & &Details \\
\hline\hline
Chassis     &  &Supermicro 6049P-E1CR36L 4U storage server\\
CPU         & 2& Intel Xeon Silver 4210R CPU \\
NIC         & 2& Mellanox MCX516A-CDAT dual-port 100GbE \\
Memory      & 8& 32 GB DDR4 memory \\
Hard drive  & 36& 16 TB Seagate Exos X18 Enterprise HDD \\
\hline 
\end{tabular}
\end{table}

\section{COSMIC Software and Data Processing Pipeline}

\begin{figure*}
    \centering
    \includegraphics[width = 0.93\textwidth]{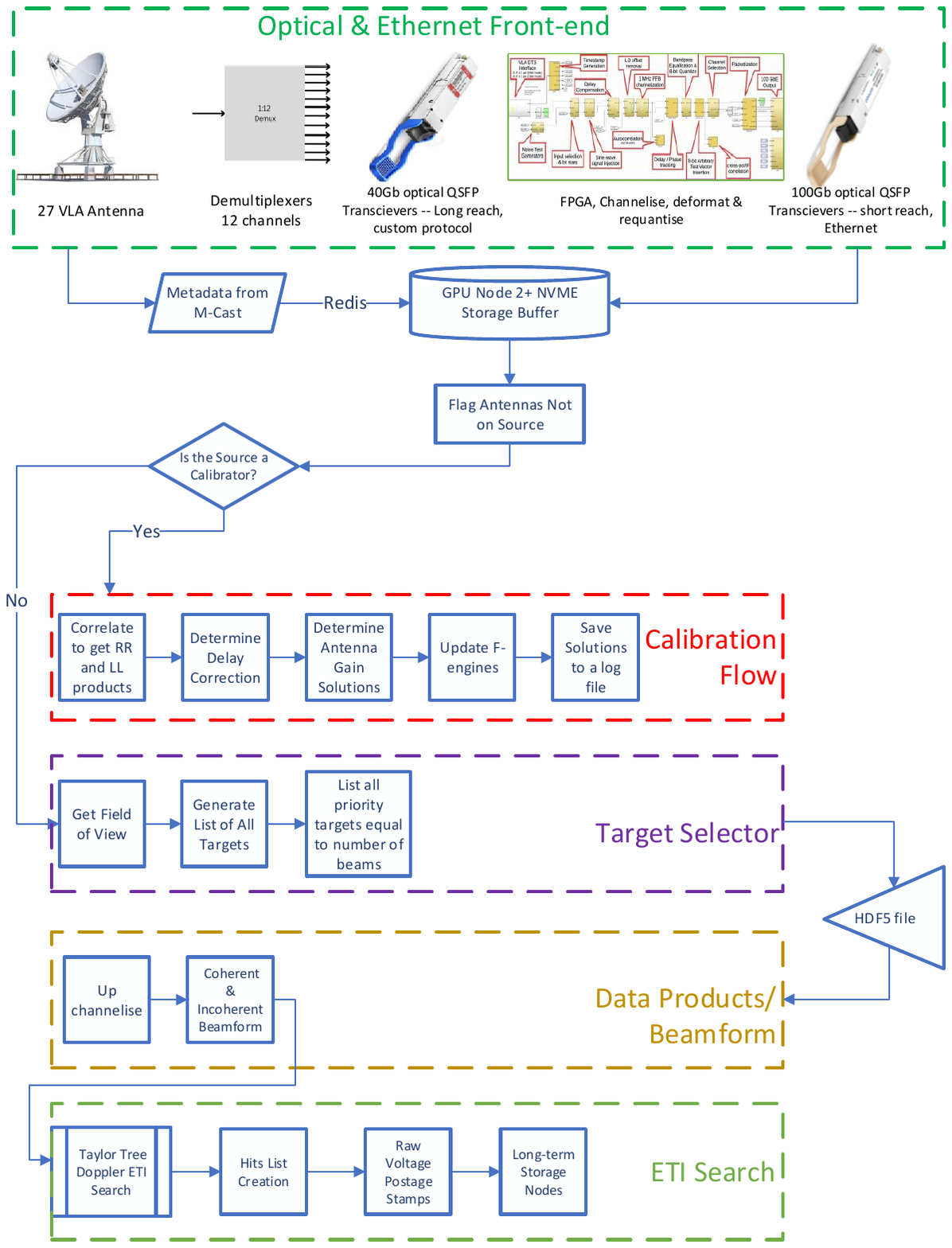}
\caption{Schematic showing an overview of the COSMIC data processing pipeline. There are two data streams: one for real-time system calibration and the other for recording and searching the data for potential technosignatures. The beamformer is designed to create a flexible number of coherent beams, depending on computing resources and workflow. Additionally, the channelization step (Upchannelizer) can be modified to account for different frequency and time resolutions.}
\vspace{-0.5em}
    \label{fig:cosmic_pipeline}
\end{figure*}

During normal operation, the NRAO broadcasts information about the telescope pointing direction, observation mode, observation frequency band, and other pertinent details regarding the scan purpose (i.e., flux calibration, phase calibration, or field) through a multicast data stream\footnote{\url{https://github.com/demorest/evla\_mcast}}, to which COSMIC subscribes\footnote{If at any time the Data Analyst at the VLA determines that COSMIC's operation would be in direct conflict with the PI's science goals, this stream is turned off such that COSMIC cannot apply meaning to the digital information. Additionally, PIs can opt out of COSMIC's commensal observation in the scheduling tool.}.

This information is read into \textsc{Redis}, an in-memory data structure store used as a distributed, in-memory key--value database, cache, and message broker. The pipeline depicted in Figure \ref{fig:cosmic_pipeline} obtains information from \textsc{Redis} about when to act on various streams in the pipeline. The data recorder uses a $YAML$\footnote{\emph{YAML Ain't Markup Language}, a data serialization language especially designed for configuration files \url{https://yaml.org/}.} as an input to allow COSMIC users to specify the criteria for the telescope state in which data recording is triggered.

\subsection{Data Recording}
Sets of 32 $\times$ 1\,MHz coarse channels are distributed across multiple downstream compute nodes in full polarization from each of the VLA antennas to perform further data processing. The 100GbE switch is used to transfer the data through the FPGA optical fiber interface to the GPU compute nodes. The data flow within the pipeline is managed using the \textsc{hashpipe} software package (an application that helps create threads and shared memory buffers between them; \citealt{MacMahon_2018}).

When the VLA is either pointing toward a target, performing on-the-fly mapping, or calibrating, \textsc{hashpipe} will capture UDP packets from the FPGA, arrange them into the Green Bank Ultimate Pulsar Processing Instrument ($GUPPI$) raw format \citep{GUPPI}, and write the data to the NVMe buffer on the GPU node. The GUPPI raw data format consists of a plain text header, loosely based on the FITS format \citep{FITS}, followed by a block of binary data. Each data block contains $2^{17}$ time samples, 32 coarse frequency channels, two polarizations, and up to 27 antennas' worth of data, amounting to an effective block size of 108\,MB. The number of time samples is chosen to ensure that each batch processing action can be performed in its most efficient configuration. However, this is flexible and could be adjusted in the future.

The metadata received from the VLA contain information on the ``intent'' of the observation, specifying whether it is a target field or a calibration source. Moreover, the VLA offers a collection of well-defined observation intents\footnote{\url{https://science.nrao.edu/facilities/vla/docs/manuals/obsguide/referencemanual-all-pages}}, which are used as criteria within the \textsc{YAML} observation configuration file. This observation configuration file specifies a pair of destination \textsc{hashpipe} instances that perform the calibration process when the observation intent is indicated as ``calibration''. The target observation ``intent'' launches a different pair of \textsc{hashpipe} instances that record the data stream into $GUPPI$ raw files. The observational \textsc{YAML} file also specifies different post-processing procedures: data collected for the calibration process are collated, and gain solutions are updated, whereas data from target observations are beamformed and searched for technosignatures. The overall data flow is shown in Figure \ref{fig:cosmic_pipeline}. Depending on the intent, the \textsc{hashpipe} automation pipeline directs the data flow to follow either the correlation and calibration pathway or the beamforming and search process.

The software across the multiple GPU servers is maintained by means of a centralized read-only operating system from which all GPU nodes boot using NetBoot\footnote{\url{https://netboot.xyz/}}, as described in \cite{MacMahon_2018}. This ensures that all of the GPU servers are running identical versions of the software and that all user access privileges are consistent across the processing systems.

\subsection{System Control}
COSMIC as a system is intended to run autonomously. \textsc{YAML} files specifying observation criteria are submitted by users to trigger the COSMIC system into action when metadata from \textsc{Redis} match those criteria. These \textsc{YAML} files contain information regarding observation frequencies and bandwidths (as these can be adjusted and do not need to incorporate the full output bandwidth by the VLA, depending on the science goals), observation intent, and scan duration. The \textsc{YAML} files also detail what sort of observation pipelines need to be engaged postrecording.

Accordingly, the only human intervention required on the recording end is the submission of observational \textsc{YAML} files to the system for monitoring.

The delay tracking and calibration process (described in \S 3.3.1) operate autonomously when the delay corrections are $<$500\,ns as expected, as in normal day-to-day operation. Delay phase tracking is chosen as the VLA updates the sky position via the multicast system. This will engage the source finder to automatically search the nearby sky, generate RA and Dec values for a point of interest, and feed this into the Delay Engine, which computes the required delays and phases for the F-Engines\footnote{`F-Engine' is a general term used to refer to the polyphase filterbank channelizer, the deformatter, and the equalization implemented on the FPGA, as shown in the fourth section within the green box in Figure \ref{fig:cosmic_pipeline} labeled ``Optical \& Ethernet Frontend''.} to apply to the relevant data with a time to load. In this way, we phase up the data streams in the F-Engines to that point of interest. This calibration process engages when a calibration target appears, triggering a correlation observation from which gains are derived, and the calibration delays and phases are produced and loaded.

If the delay corrections are greater than 500\,ns, such as when the antennas move to a new configuration, human intervention is needed. In this scenario, an observation for calibration and cross-correlation will not yield the frequency resolution required to produce new delay values. Therefore, it is necessary to submit a special \textsc{YAML} file that instructs the \textsc{hashpipe} instance to record an observation of a bright calibrator source and save the $GUPPI$ raw files without searching and then to use a bespoke \textsc{Python} correlator to upchannelize and produce calibration delay values. We then manually inspect the generated delays and submit them as the new default for the calibration process, if they are deemed appropriate. If the values are not appropriate, the data will be further investigated, and the system will not be used for science until new appropriate values are obtained.

\subsection{Correlation and Calibration}
When the VLA is pointed at, and the observer has marked the observation intent as, a flux density, bandpass, phase, or gain calibrator source, the raw voltages from each antenna are cross-correlated using the \textsc{xGPU}\footnote{\url{https://github.com/GPU-correlators/xGPU/blob/master/README}} \citep{Clark:2013} software correlator to produce visibility data products in four polarizations (RR, LL, RL and LR, where R and L refer to right and left circular polarization, respectively). The software correlator writes these data products into a data file in the $UVH5$ format\footnote{\url{https://pyuvdata.readthedocs.io/en/v1.5/_modules/pyuvdata/uvh5.html}}, which contains all of the information required for real-time calibration but can be used in imaging applications as well.

For calibration, we follow standard calibration procedures utilized for interferometric telescopes, but these procedures are executed through a real-time and autonomous process. The NRAO provides detailed information about VLA calibration on their website for users of the telescope\footnote{\url {https://science.nrao.edu/facilities/vla/docs/manuals/obsguide/calibration}}, which we follow for the general purpose of calibration. For technosignature detection, the primary motivation for COSMIC, two types of real-time calibration are implemented; delay calibration and gain calibration. Currently, we are not conducting any bandpass or amplitude calibrations in real time, although the correlated data of amplitude and bandpass calibrators are saved in case we need to calibrate the bandpass responses and amplitudes following the detection of of extraterrestrial intelligence (ETI). However, amplitude equalization is conducted in the FPGAs at the time of configuration to normalize the signal levels from all antennas. This is done to ensure that the beamformed output is not dominated by signals from certain antennas. We have determined that through this process, we obtain an equivalent system flux density within 10\% of the value reported by the NRAO for the VLA.

\subsubsection{Delay Calibration}
As discussed in \S 2, COSMIC receives a copy of the digitized voltages from the antennas along a signal pathway after the splitter that is different from that of the existing VLA \textsc{WIDAR} system.
The total delay in the signal chain a the sum of the fixed (instrumental, nongeometric) delays and geometric delays. Testing of the COSMIC system with the \textsc{WIDAR} delays failed to produce coherence, as \textsc{WIDAR} uses different time stamping methods and the cable lengths are slightly different between COSMIC and \textsc{WIDAR}. Therefore, a bespoke delay model was created (\S \ref{sssec:delaymodel}), and COSMIC-specific fixed delays are calculated after every antenna configuration change (see \ref{sssec:fixeddelays}). The delay calibration consists of two steps. The first step is the estimation of the fixed delays, which are mostly associated with the fibers, electronics, etc. The second is the estimation of the geometric delays, which can be easily calculated using accurate models. The fixed delays should be roughly constant over time, and we expect up to 10--15\,ns differences for each of the VLA receivers. In contrast, the geometric delays will change as functions of time, baseline, and source direction.

\subsubsection{Fixed Delays}
\label{sssec:fixeddelays}
A recording of a bright calibrator is taken with zero delays applied in the signal chain. The total delay is calculated as the inverse fast Fourier transform (FFT) of the cross-correlated integrated spectra for each baseline (pair of antennas) to produce an associated per-baseline delay peak. This delay is translated into a per-antenna delay by selecting a suitable reference antenna (i.e., one close to the center of the array) that is known to be recorded correctly. Geometric delays are calculated retroactively for each observation (based on source pointing, antenna position, and time) and subtracted from this total delay. This leaves a per-antenna fixed delay (calibration delay) that serves as the new constant delay offset for observations in the current array configuration\footnote{For more information on the VLA configurations, see \url{https://public.nrao.edu/vla-configurations/}.}.

These fixed delays are measured during the start of each VLA reconfiguration and are updated in the FPGAs accordingly (see \S 3.2 for a detailed explanation). During the commissioning phase of COSMIC and while the VLA was in the C configuration, observations of bright calibrators spanning multiple weeks were conducted to ensure the consistency of the fixed delay values as functions of time, pointing, and frequency. The results showed consistent phase correction and calibration.

A sum of the fixed and geometric delays based on the VLA model (see \S 3.3.2) is used to compensate each antenna in the F-Engines in real time. This ensures that the delay tracking of sources in the observation is correct. We have implemented all of the delay tracking functions as software operations acting on the recorded telescope raw voltages.

\subsubsection{Delay Tracking and the Geometric Delay Model}
\label{sssec:delaymodel}
As the Earth rotates, the projection of each baseline changes with respect to the phase tracking center. The phase tracking centers (the placement of the coherent beams) are chosen by the target selector (\S 3.5) to be within the full width at half maximum (FWHM) of the primary beam.

The phase-centered controller can operate in two modes. In mode 1, the RA and Dec values are sent out at specific intervals, usually corresponding to updates from the multicast system through \textsc{Redis}, and the delay engine is left to derive phase solutions for the current pointing location. In mode 2, at time intervals that align with the $GUPPI$ raw file boundaries (8 seconds for the B, C, and D configurations of the VLA and 2 seconds for the A configuration) during data recording, the phase-center controller updates the phase pointing such that each file is phased to only a single pointing location. This precision is achieved by associating load times ($t_\mathrm{{load}}$) with the pointing coordinates and sending the pointing coordinates and load times out several seconds before $t_\mathrm{{load}}$.

The delay model receives the pointing positions and associated $t_\mathrm{{load}}$ values via a \textsc{Redis} channel and, using the positions of the antennas along with the time of reception $t_\mathrm{r}$, derives quadratic coefficients $C_0$, $C_1$ and $C_2$ such that the delay at $t_\mathrm{r}$ is calculated as follows:

\begin{equation}
    Delay(t_\mathrm{r}) = C_0 + C_1\times t_\mathrm{r} + \frac{C_2}{2}\times t_\mathrm{r}^{2}
\end{equation}

The delay model then sends the coefficients $C_0, C_1$, and $C_2$ out to each of the F-Engines via separate \textsc{Redis} channels, along with the current scans of $sideband$ and $center$ $frequency$, $t_\mathrm{r}$, and the received $t_\mathrm{{load}}$.

Finally, the F-Engines receive this information, with which they perform a quadratic interpolation to the provided $t_\mathrm{{load}}$ for both the delay

\begin{equation}
    Delay(t_\mathrm{{load}}) = \tau_s + C_0 + C_1\times t_\mathrm{{diff}} + \frac{C_2}{2}\times t_\mathrm{{diff}}^2
\end{equation}

\noindent and the delay rate

\begin{equation}
    \frac{d}{dt}(Delay(t_\mathrm{{load}})) = C_1 + C_2\times t_\mathrm{{diff}}^2
\end{equation}

\noindent where

\begin{equation}
    t_\mathrm{{diff}} = t_\mathrm{{load}} - t_\mathrm{r}
\end{equation}

\noindent and $\tau_s$ is the fixed calibration delay derived from a $GUPPI$ raw file for the FPGA streams\footnote{A stream is defined as an LO tuning (IF) for a single polarization; therefore, COSMIC has 4 streams of incoming data.} in seconds ($s$).

$Delay(t_\mathrm{{load}})$ has units of nanoseconds (ns) and is a decimal float value. The integer part is separated and applied as a coarse delay (prechannelization), while the fractional part is applied as a fine delay (postchannelization).
$\frac{d}{dt}(Delay(t_\mathrm{{load}}))$ has units of $ns\,s^{-1}$ and is also loaded into the F-Engines for the FPGAs to perform linear interpolation from the loaded fractional delay.

It is imperative to continuously update the delays on the F-Engines, as linear interpolation does not provide the accuracy required to track sources accurately and the fractional delays in the F-Engines are susceptible to information overflow. For this reason, for each antenna on each F-Engine node, a \textsc{Python} thread is spawned at 0.5\,s intervals to quadratically reinterpolate the delay and delay rates to $t_\mathrm{{load}}$ (if no specific $t_\mathrm{{load}}$ is provided, $t_\mathrm{{load}}$ is set to 0.5\,s into the future) and load the results into the F-Engines at $t_\mathrm{{load}}$.

Simultaneously, a phase and phase rate are derived from the computed delay and delay rate, which are also applied postchannelization, as follows:

\begin{equation}
    \Phi(t_{load}) = 2\pi\times lo_{cf} \times sb_s \times lo_{cf} \times sslo_s \times delay(t_{load})
\end{equation}

\noindent and

\begin{equation}
    \frac{d}{dt}(\Phi(t_{load})) =  2\pi \times sb_s \times sslo_s \times \frac{d}{dt}(delay(t_{load}))
\end{equation}

\noindent where $sb_s$ is the per-stream sideband value, $sslo_s$ is the per-stream effective LO, and
$lo_{cf} = 2\pi \times sb_s * f_{shift}$ is a phase-correction factor that needs to be applied due to the $f_\mathrm{{shift}}$ mixing performed for each FPGA stream before fine channelization.

\subsubsection{Gain Calibration}
The delay errors are treated as fixed values when generating the output of the calibration pipeline during delay tracking, but due to uncertainties (of less than 10\,ns) in the fixed delay values, there are still some residual uncorrected delays. The first priority of COSMIC is the detection of technosignature signals, and the only calibration that we consider after delay calibration is the calibration of the complex gain; no absolute flux density calibration is performed.

Gain calibration is an antenna-based correction that accounts for time-varying factors associated with the instrument and the atmosphere. If these factors remain uncorrected, the differences in the gains for each antenna can impact the phasing of the incoming voltage streams, leading to decorrelation of the sum of the antenna signals.

To conduct the gain calibration, we use a modified version of the \textsc{sdmpy}\footnote{\url{https://github.com/demorest/sdmpy/tree/master}} package, which was initially written to calibrate the VLA Science Data Model (SDM) data sets for the $realfast$ commensal system \citep{Realfast}. In this version of the package, the calibration utility in the \textsc{sdmpy} software has been modified to work with the COSMIC $UVH5$ data format.

The gain calibration solutions for each stream are written out as a {\tt Python} dictionary. The \textit{derive\_gains} method in the \textit{calibrate\_uvh5} class of \textit{calibrate\_uvh5.py}\footnote{\url{https://github.com/COSMIC-SETI/COSMIC-VLA-CalibrationEngine/blob/rfi\_mitigation\_and\_arrayconfig\_update/calibrate\_uvh5.py}}
is used to derive the antenna gains. Gain calibration is carried out independently on each 32\,MHz bandwidth (32$\times$1\,MHz coarse channels) of the correlated $UVH5$-formatted data (for each LO tuning), which are distributed across multiple GPU compute nodes. This is done such that each of the 44 compute nodes (2 compute nodes per GPU server) has 32\,MHz of data to process. The resultant gain dictionaries are sent to the head node, where the gains across the multiple 32\,MHz subbands are combined along the frequency axis for further processing. After all gain dictionaries have been collected for the 1\,MHz coarse channels for both LO tunings, all polarizations, and each antenna, the head node sorts and orders the complex gains into a \textsc{Python} {\tt numpy} matrix with dimensions of $n_\mathrm{{pol}} \times n_\mathrm{{tunings}}$ and $n_\mathrm{{frequencies}}$.

This resulting gain matrix is fed into a calibration kernel that can perform either a linear or Fourier interpolation to calculate the residual delays and phases for each antenna and each polarization. The calibration phases previously loaded into the FPGAs are also subtracted from the received gain matrix to calculate the new delay residuals and phases that are provided to the F-Engines.

A plot of the residual delays and the per-antenna phase solutions for both IFs are also uploaded to a \textsc{Slack}\footnote{\url{https://slack.com/}} interface channel in real time to allow scrutiny by all collaborating scientists. The amplitude plots are also uploaded for review even though no amplitude calibration is performed.

\begin{figure*}[h]
    \centering
    \includegraphics[width = 0.95\textwidth]{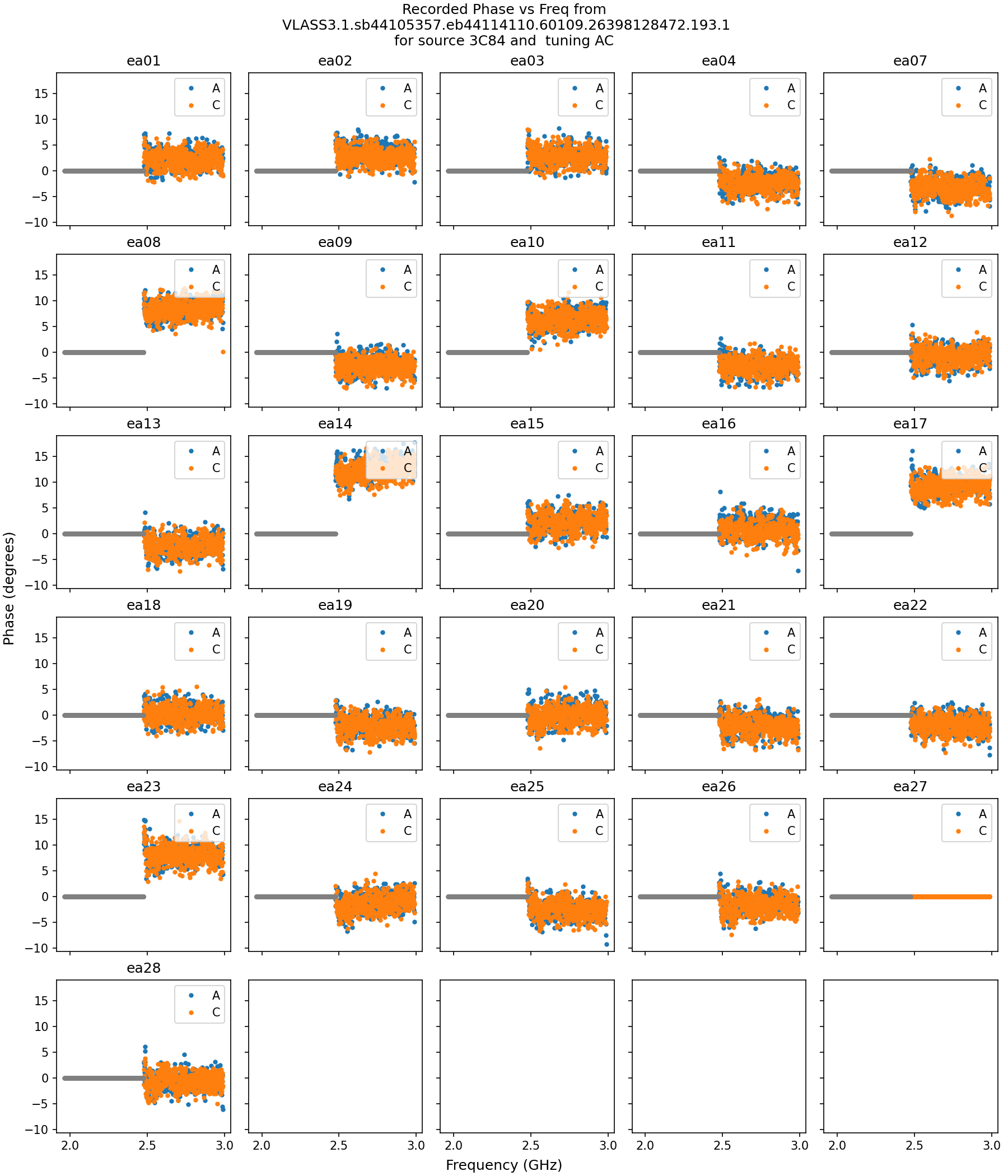}
\caption{The phases of the gain solutions recorded for each data set and plotted as a function of frequency for each antenna in both polarizations of the AC tuning (IF). Here, AC refers to a signal pair tuned to the same frequency, where A is the right circular polarization and C is the left circular polarization\footnote{\url{https://science.nrao.edu/facilities/vla/docs/manuals/oss/performance/vla-samplers}}. Each tuning collects 512\,MHz of the total bandwidth of 1024\,MHz. For all data with no recorded phase values, a value of zero is set as a placeholder.}
    \label{fig:gain_phase}
\end{figure*}

Figure \ref{fig:gain_phase} shows an example plot of the phases of the gain solutions plotted as a function of frequency for each antenna that is sent to the designated \textsc{Slack} channel for a single calibration observation. The current COSMIC system collects data only in the second half of each IF, spanning 512\,MHz\footnote{At the time of these observations, only 512\,MHz of bandwidth was recorded for each LO. However, during the writing of this paper, we upgraded to 22 compute nodes, and we can therefore now record up to 704\,MHz bandwidth for each IF. Our explanations throughout the rest of this paper will continue to refer to what was in place during science commissioning.}, due to computational resource limitations, which is why some of the band is grayed out and zero (see \S 5.1 for a discussion of upgrades to the recording bandwidth). The plot shows consistent flat phases close to zero for each operational antenna, as expected for well-behaved phase calibration. We note that radio frequency interference (RFI) from sources external or internal to the antenna data stream can introduce bad phase solutions, especially near the region of 1--2\,GHz, which is heavily affected by satellite RFI.

\subsection{Evaluation of Calibration}
Along with the real-time calibration, we also calculate certain statistical parameters to determine the quality of the calibration.
\begin{itemize}
\item The inverse FFT of the correlated spectra for each antenna with respect to a reference antenna is used to estimate the residual delay. The signal-to-noise ratio (SNR) of the delay peak serves as a measure of the coherence between voltages for the corresponding baseline. A low SNR value for a baseline indicates decoherence between the voltages. If the SNR value is 4 or less, then the calibration is not applied to the F-Engines.
\item The standard deviation of the phases is calculated to understand their spread.
\item An antenna grade\\
\begin{equation}
        G_{ant} = \frac{|\sum_{\nu = 0}^{\nu = n\_chan} Gains(\nu, ant)|}{ \sum_{\nu = 0}^{\nu = n\_chan} |Gains(\nu, ant)|}
    \end{equation} \\
is calculated for each stream. A value of $G_{ant} = 1$ indicates a flat zero phase across frequencies and, thus, a good calibration. In contrast, $G_{ant} = 0$ results from a large variation in phase across frequencies, which indicates a poor calibration.

\item A frequency channel grade \\
\begin{equation}
        G_{\nu} = 
    \frac{|\sum_{ant = 0}^{ant = n\_ant} Gains(\nu, ant)|}{ \sum_{ant = 0}^{ant = n\_ant} |Gains(\nu, ant)|}
    \end{equation} \\
is calculated for each stream. A value of $G_{\nu} = 1$ indicates that the antennas are properly phased to the intended position, implying a good calibration. $G_{\nu} = 0 $ indicates phase differences between the antennas resulting from a poor calibration.

\item An overall grade \\
\begin{equation}
         G = \frac{\left | \sum_{ant,\nu = 0}^{ant,\nu = n\_ant,n\_s,n\_chan} Gains(\nu,ant,s) \right |} {\sum_{ant,\nu = 0}^{ant,\nu = n\_ant,n\_s,n\_chan}|Gains(\nu,ant,s)|}
    \end{equation}
is calculated across the full gain matrix. This overall grade tends toward unity when the calibration recording is phased to the intended position across all antennas, streams, and frequencies. The overall grade tends toward zero when the calibration recording is incorrectly phased across one or more of these variables.
\end{itemize}

\subsection{Antenna Flagging and RFI Mitigation}
The start of VLA observations is triggered when the antennas start to slew toward the designated source. However, the antenna information received through \textsc{Redis} will show that the antennas are not ``on source'', and therefore they will be flagged accordingly, until they arrive at the target location. Additionally, any antennas that are offline (stowed) and therefore never ``on source'' are flagged prior to correlation or raw data recording. The COSMIC system also flags any antenna that has unstable DTS information and thus does not record data from antennas with variable power.

RFI consists of human-made signals not related to the signal being sought, often generated by terrestrial technology or satellites. In our search for technosignatures, we aim to eliminate RFI---both external and intrinsic to the processing pipeline---to seek out signals associated with extraterrestrial technology. Although RFI detection and excision is a critical step in determining the difference between signals of terrestrial and extraterrestrial origin, the current COSMIC pipeline does not implement real-time RFI excision; therefore, early data can be used to evaluate the RFI environment and to experiment with potential excision algorithms on representative data. Over the many years of operation of the VLA, the observatory has gathered a vast amount of data regarding the RFI problem. Therefore, when we initially evaluate the data for scientific merit, we rely on these historical data\footnote{\url{https://science.nrao.edu/facilities/vla/docs/manuals/obsguide/rfi}} from the facility to avoid the regions of the observation band that are significantly affected by RFI.

\begin{figure}
    \centering
    \includegraphics[width = 0.48\textwidth]{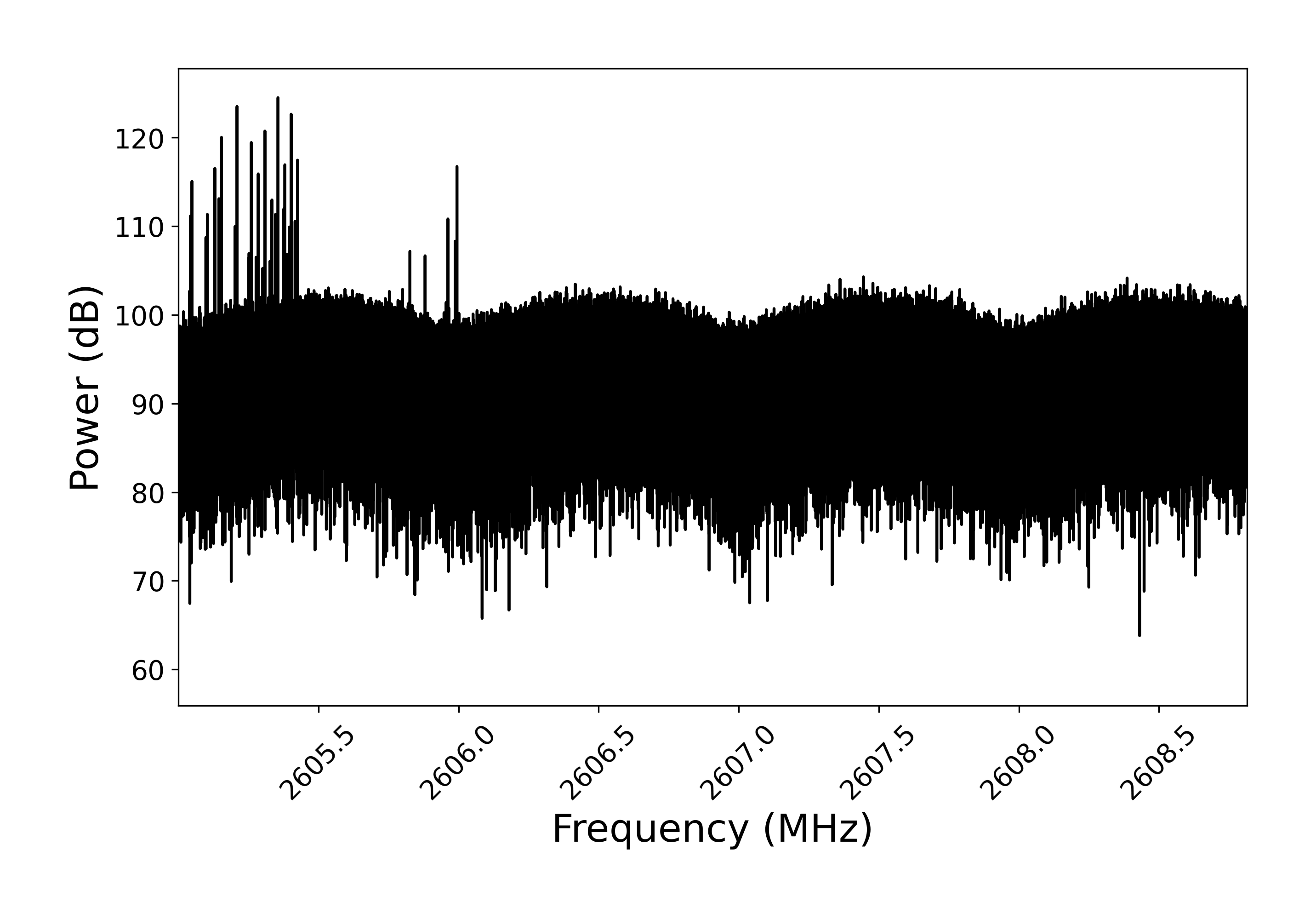}
    \includegraphics[width = 0.48\textwidth]{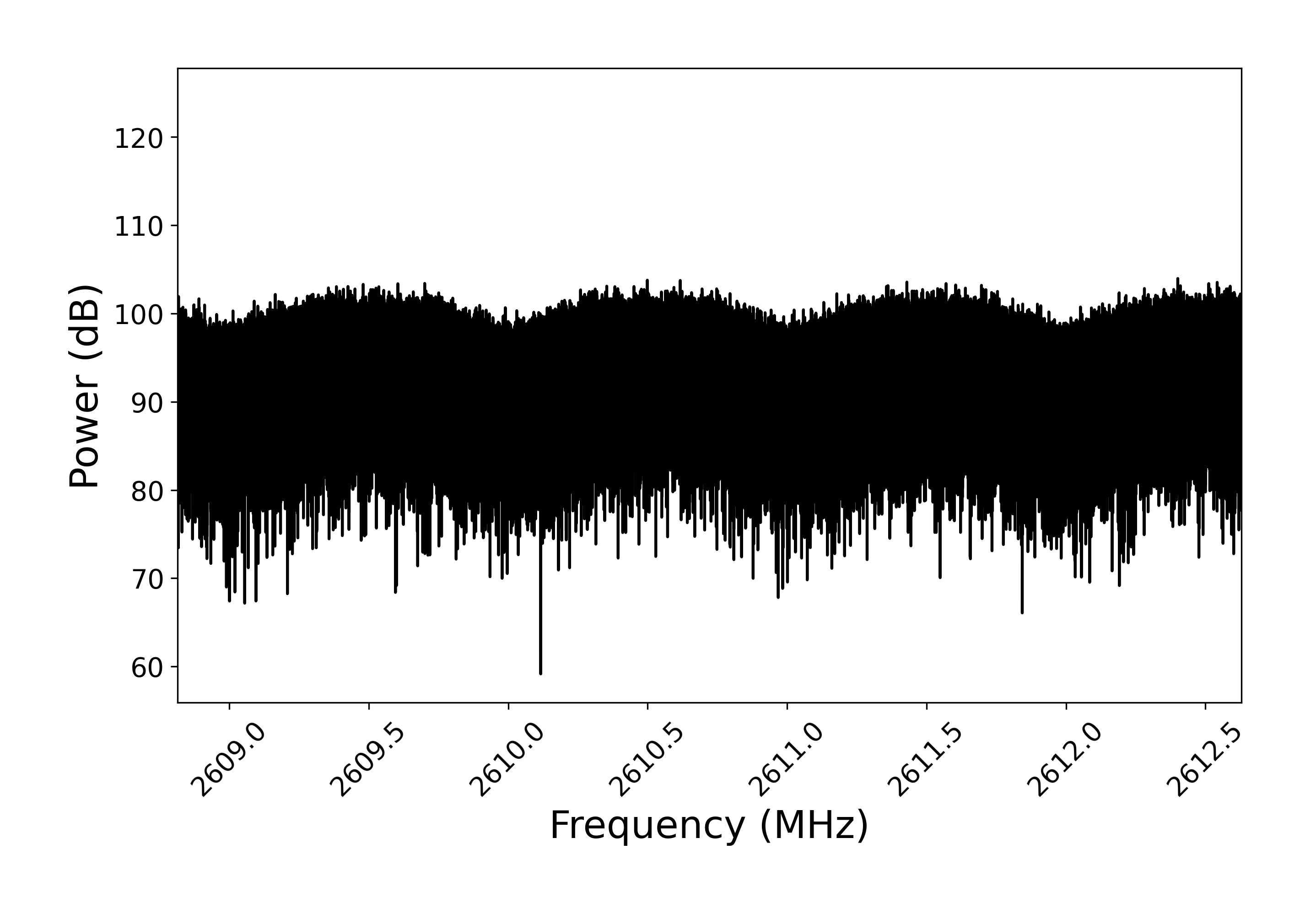}
\caption{An example of the power spectra within the 2--3\,GHz band. The sharp spikes are persistent narrow-band RFI, whereas most of the band is free of such features. The ``clamshell'' shape arises from bandpass filtering using the polyphase filterbank. The channel resolution in these spectra is 8\,Hz.}
\vspace{-0.5em}
    \label{fig:Power_spectra}
\end{figure}

We note that the RFI environment for extremely high-resolution spectroscopy, such as the technosignature data of interest to COSMIC, can be much more complex than the environment for the signals typically acquired for general observatory purposes. It is envisioned that in the future, a database of RFI will be created within the COSMIC data processing pipeline to enable us to mask these interfering signals before beginning a drift search. However, we can presently identify regions of strong RFI in which signals should be disregarded on a first pass through the data. An example of power spectra from one night of observation is shown in Figure \ref{fig:Power_spectra}.

\subsection{Target Selection}
The NRAO's multicast system continuously provides the current pointing for each observation, from which the target selector software\footnote{\url{https://github.com/danielczech/targets-minimal}} calculates the primary field of view (FoV).\footnote{For the FWHM power in arcminutes of the FoV, we use the NRAO-suggested formula of 42/$\nu$\,GHz for frequencies between 1 and 50\,GHz, and at 700\,MHz, we use an approximate value of 50/$\nu$\,GHz.} (see Figure~\ref{fig:vlass_obs}). Using this FoV, the target selector determines which stars are available for observation and calculates an optimized list of the coordinates of the highest interest at which to form coherent beams. The target selector draws these target stars and other objects (falling within the primary FoV) from several databases, including a 32 million star sample derived from $Gaia$ Data Release II \citep{Czech_2021} and the Breakthrough Listen Exotica Catalog \citep{Lacki_2021}. The targets are ranked in priority based on their distance. The goal is to search as many unique stars as possible, prioritizing our closest neighbors. The target coordinates are fed into a special $HDF5$-based ``beamformer recipe file ($BFR5$)'' used by the beamformer to form coherent beams.

\subsection{Coherent and Incoherent Beamforming}
COSMIC forms at least five coherent tracking beams, depending on the available computing resources and the amount of time spent recording. The idea is to form all of the beams at once over the full recorded FoV. The raw voltage data are stored in $GUPPI$ raw binary files, which contain blocks of data and header information.

All of the time samples in a $GUPPI$ raw file are acquired to perform an FFT to finely channelize the recorded data stream from the 1\,MHz coarse channels to $\sim$8\,Hz fine frequency channels. The beamforming phase coefficients are calculated as follows:

\begin{equation}
    \textbf{w}(\phi) = e^{j\phi}
\end{equation}

\noindent
where
\begin{equation}
    \phi = 2\pi f \tau + \psi_{f}
\end{equation}
\noindent
$f$ is the center frequency of the coarse (1\,MHz) channel, $\tau$ is the delay relative to the reference antenna, and $\psi_{f}$ is the phase calibration solution for a particular coarse channel.

The $BFR5$ ``beamformer recipe file'' provides the list of target source coordinates and other relevant information to inform the calculation of the phase coefficients for the beam formation process, which is performed using the Breakthrough Listen Accelerated DSP Engine (\textsc{BLADE}; L. Cruz et al (in prep)\footnote{\url{https://github.com/luigifcruz/blade}}), a software suite designed for the Allen Telescope Array in California, USA, and modified for use in the COSMIC pipeline. Currently, COSMIC performs a 131027-point FFT and forms five coherent beams plus an incoherent beam. The resulting total of 64 time samples per spectrum is channelized to 7.6294\,Hz given a 1\,MHz coarse channel. Only one set of phase coefficients is used for beam formation for the recorded time span of 8.388\,s (when the VLA is in the B, C, or D configuration) or $\sim$2\,s (when the VLA is in the A configuration). However, these values can be flexibly tuned depending on the situation. This approach results in an average beamforming computation time on the GPU of approximately 50\,s from file ingestion to the generation of output from \textsc{BLADE} (including memory transfers). Improvements to the \textsc{BLADE} software package to decrease this processing time are ongoing.

At the output of the beamformer, the raw voltages of the coherent beams are converted into power, and the polarizations are summed to a pseudo-Stokes I total intensity. These processes are all implemented on the GPU with $CUDA$. The output is then passed to the technosignature search algorithm discussed in \S 3.6.

To assess the performance of the beamformer and ensure that the phases and delays are appropriately accounted for as well as to verify the general operation of the beamformer, a bright Class II methanol maser was observed at 6.7\,GHz. We obtained test observations of W51M, which, according to \cite{Etoka_2012}, has a brightness of $\sim$250\,Jy and is close to a point source at the VLA B-configuration\footnote{\url{https://science.nrao.edu/facilities/vla/docs/manuals/oss/performance/resolution}} resolution of $\sim$2\,arcseconds. We imaged the VLA \textsc{WIDAR}-collected data using \textsc{CASA} \citep{CASA} to find the exact location of the source and independently verify the topocentric frequency of emission in comparison to COSMIC, as both data sets were recorded simultaneously. We determined the source position to lie at RA(J2000) = 19$^{h}$23$^{m}$43.95$^{s}$ and Dec(J2000) = 14$^{\circ}$30$^{m}$34.34$^{s}$, with the topocentric frequency of the peak emission being at 6667.986\,MHz. By looking at the time-averaged power spectrum of the coherent beamformed data from COSMIC, we found that the frequencies matched and that the signal was detected at a velocity that was well matched with the published results. Thus, by forming coherent beams at a 0.25\,arcsecond separation around the \textsc{WIDAR} position for the source, we achieved a pointing accuracy within half of the VLA point spread function.

\begin{figure}
    \includegraphics[width = 0.49\textwidth]{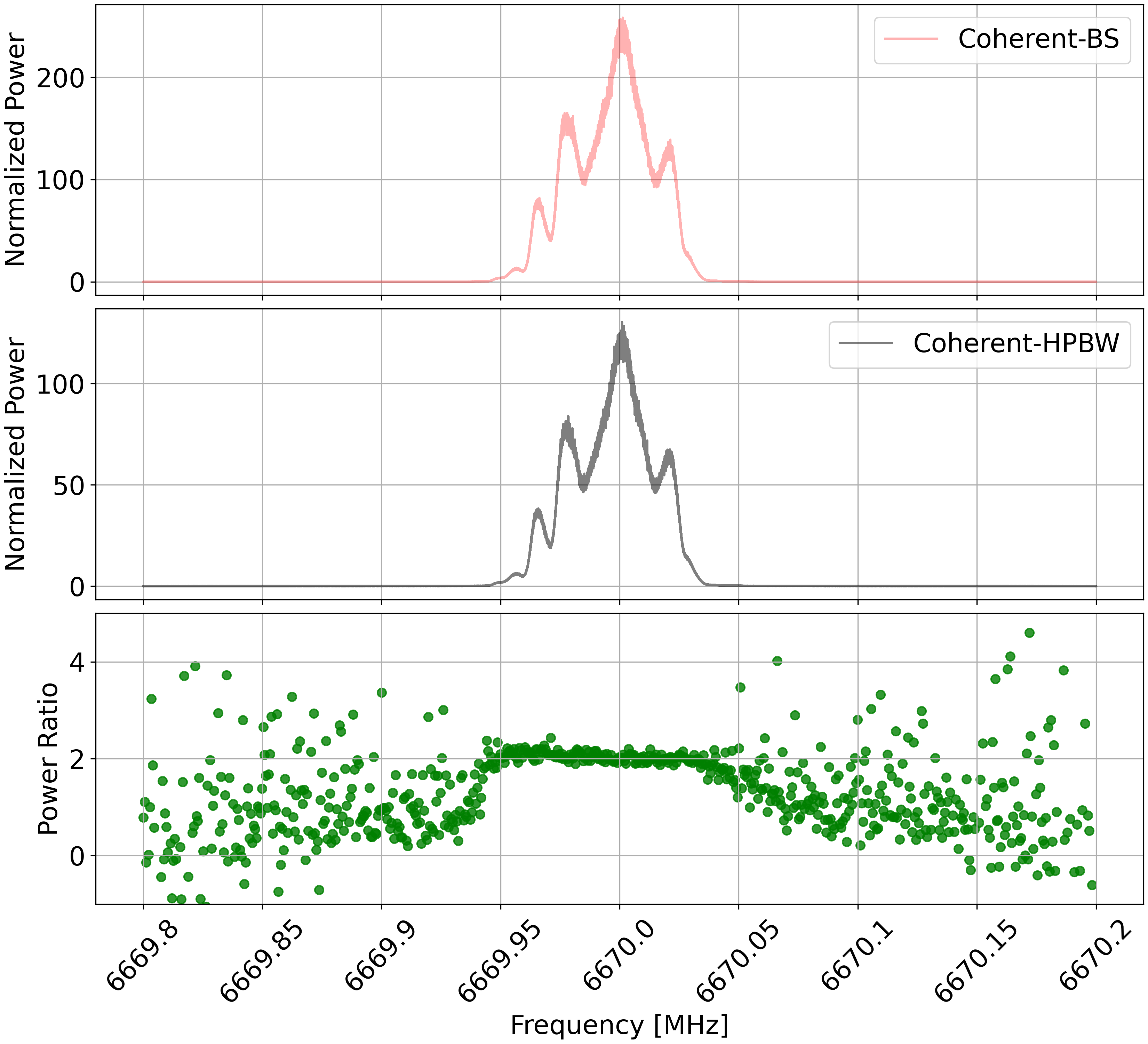}
\caption{Spectra from the W3OH methanol maser source when observed at the boresight (BS; red) and at the half-power point of the primary beam (HPBW; black), both phasing toward the peak position of the maser source as determined from an image generated with data from \textsc{WIDAR}.}
\vspace{-0.5em}
    \label{fig:halfpower}
\end{figure}

To test the sensitivity of the expected flux density recovery, with the VLA in the A configuration, we obtained two observations of the Class II methanol maser associated with W3OH at RA(J2000) = 02$^{h}$27$^{m}$03.8192$^{s}$ and Dec(J2000) = 61$^{\circ}$52$^{m}$25.230$^{s}$. In one observation, the source was near the center of the primary beam (boresight), and it was at the half-power point of the primary beam in the other, approximately 2\,arcminutes from the phase center for the A configuration, each with $\sim$8\,s of recorded data from the source. These two observations were recorded with 22 operational VLA antennas (the data were taken during maintenance, so the full array was not available), while COSMIC was configured with the raw data output and save mode turned on but the technosignature search turned off. As observed in Figure \ref{fig:halfpower}, the maser at the FWHM of the primary beam exhibits 1/2 the power level of the maser at the boresight (a $\sim$2\,dB difference), as expected.

\subsubsection{Beamformer Efficiency}
To measure the beamformer efficiency, observations of how the SNR changes with the addition of antennas were computed as an overall measure of the correlated sky noise \citep{Kudale_2017}. For this analysis, we formed coherent beams toward the W51M methanol maser, a strong emitter for which perturbations in the SNR can be reduced by using the frequency channel with the most intense signal and the time-averaged power over a five-minute observation. The observations were calibrated as described in \S3, with an evaluation of the phase stability before and after the maser observation. Using the \textsc {BLADE} beamforming code, we incrementally added 3 to 21 antennas in sets of 3 antennas in each beamforming round. The noise was calculated by taking the mean of 4 000 RFI-free channels (each 8\,Hz wide, for a total of 32\,kHz) from the same 1\,MHz coarse channel as the maser emission line but not including the maser emission. The SNR was then calculated by subtracting the peak power of the source and the noise, both in units of dB.

\begin{figure}
    \centering
    \includegraphics[width = 0.48\textwidth]{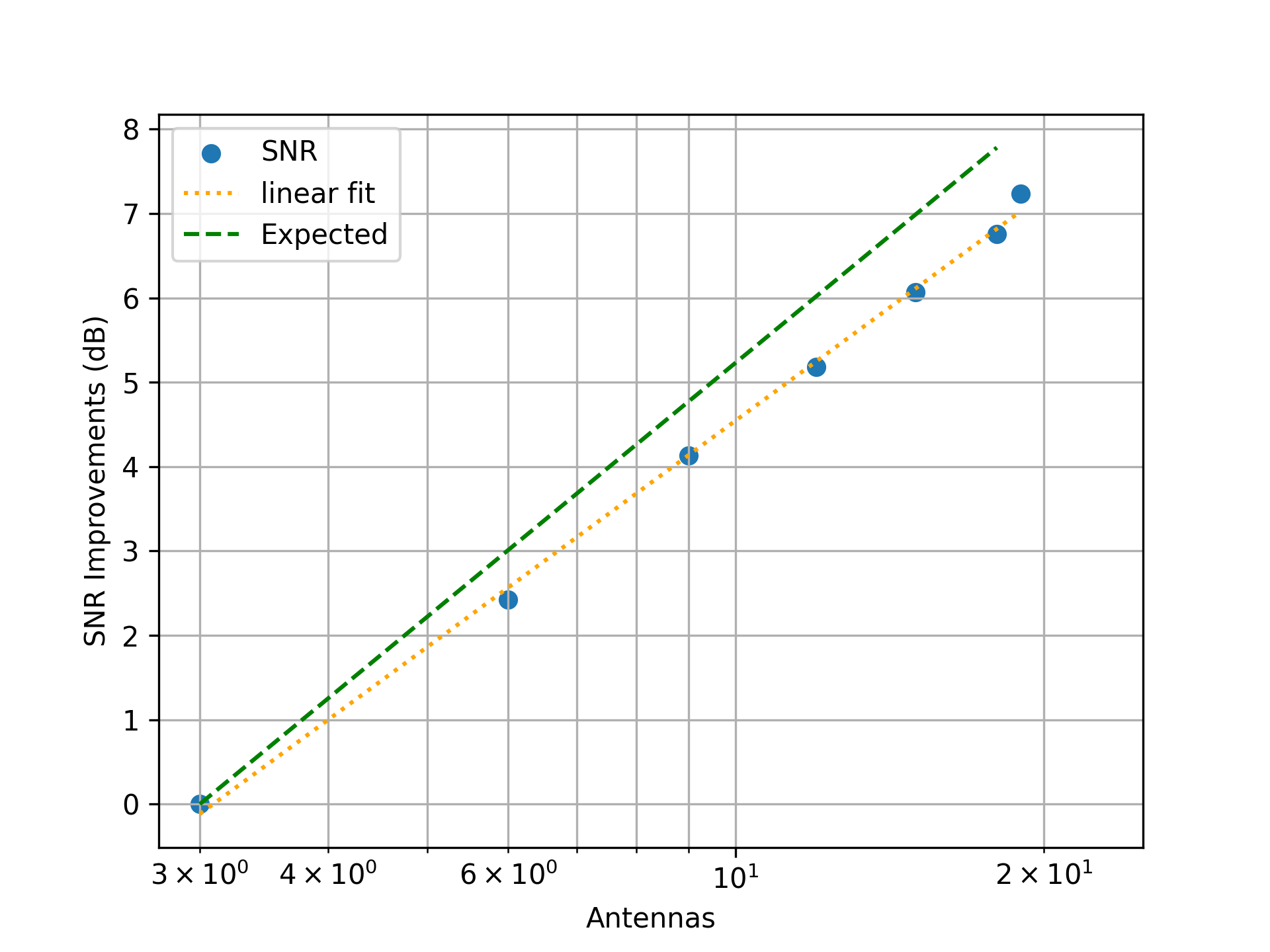}
\caption{SNR improvement when adding more antennas.}
\vspace{-0.5em}
    \label{fig:efficiency}
\end{figure}

To determine the SNR improvement, which is shown on the y-axis in Figure \ref{fig:efficiency}, all values were then subtracted from the value obtained from only three antennas (the first data point). The ideal SNR improvement is calculated as a linear increase from 0 to 7.8\,dB (10*log10(18) - 10*log10(3) = 7.8 dB), representing the theoretical expectation for a 100\,percent operational system. The plot shows that as more antennas are sequentially added, the SNR increases linearly, as expected. However, the total power is approximately 20\% lower than expected if all signal paths for all antennas are functioning at the theoretical total efficiency, with some contribution from phase calibration errors. This is within the expected tolerances for the system.

\subsubsection{Coherent vs. Incoherent Beams}
A coherent beam, in the receiving paradigm, is produced by altering the phase on the elements of an array such that parallel-plane electromagnetic radiation emanating from a particular source combine constructively (\S 3.7). This method trades off beam size for sensitivity when compared to an incoherent beam. An incoherent beam, produced by summing the total intensity of all receiving elements (disregarding phase information by definition) exposes sources across the FoV of each telescope, which is still important when performing a blind SETI search.


We thus compared the coherent and incoherent beam formation processes to ensure that the expected decrease in sensitivity for the W51M class II methanol maser signal was observed. The beamforming process was conducted on the same observation with the same integration time and the same number of antennas, with a phase calibration completed as described in \S3.3. Because the data were not flux density calibrated, we evaluated the ratio based on the normalized power of the two signal strengths for the peak emission from the source.

For the incoherent summation, the noise ($\sigma$), the source flux density ($S_{src}$) and the equivalent system flux density ($S_{sys}$) are related as follows:
\begin{enumerate}
\item $S_{src} / S_{sys}$ is constant.
\item $S_{src} / \sigma$ scales as $\sqrt{N}$, where $N$ is the number of antennas.
\end{enumerate}

For the coherent summation, the following relations hold:
\begin{enumerate}
\item $S_{src}  / S_{sys}$ scales as N.
\item $S_{src} / \sigma$ also scales as N.
\end{enumerate}

Therefore, if the SNR is defined as $S_{src} / \sigma$, the coherent and incoherent summations on the same data set can be related by

\begin{equation}
    Norm\,Power_{\mathrm{coherent}} / Norm\,Power_{\mathrm{incoherent}} = N
\end{equation}

\begin{figure}
    \centering
    \includegraphics[width = 0.49\textwidth]{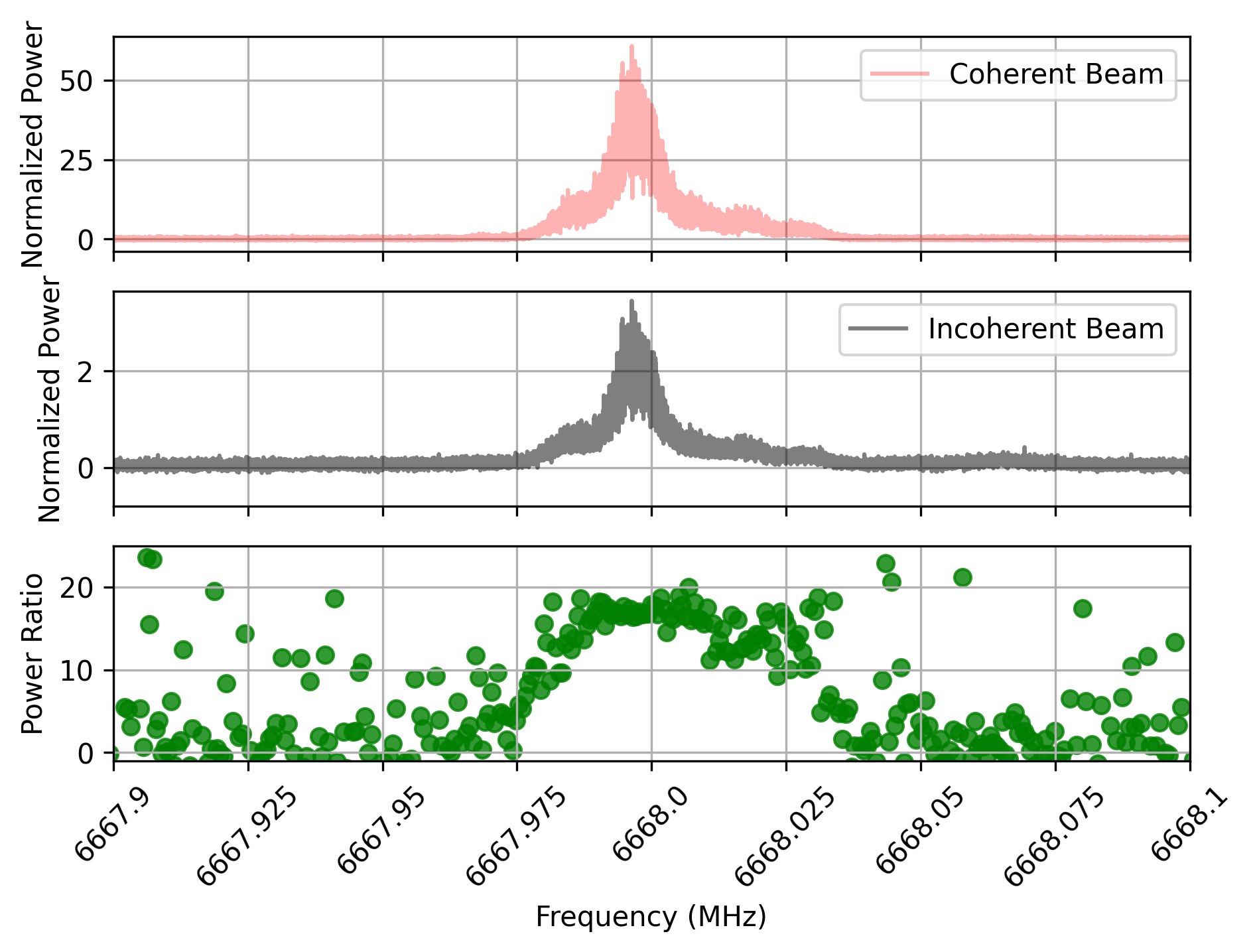}
\caption{Methanol maser emission from W51M as observed in a coherent beam (red) and an incoherent beam (black). The power ratio is approximately 11\,\% lower than the expected ratio of 23 due to calibration errors. The scatter on the edges arises because the calculation involves taking a ratio of very small numbers.}
\vspace{-0.5em}
    \label{fig:Incohernt}
\end{figure}

Here, the $Norm\,Power$ value is calculated by subtracting the power for a frequency range that is far from the source of the signal from the power for the spectral frequency range of the signal and dividing by the off-source power ((on -- off)/off). For an observation of the W51M methanol maser, we performed coherent and incoherent beamforming operations on the same data set with 23 operational antennas. As shown in Figure \ref{fig:Incohernt}, we found a maximum ratio of 20.5, which suggests an 89\,percent efficiency, similar to the value computed in Figure \ref{fig:efficiency}. As seen from the phase calibration plots for this observation, four antennas had problems with their phase delays, contributing to this reduced efficiency. This is considered an acceptable level for an autonomous real-time system, but further improvements will be sought. 

\subsection{Technosignature Search}
Hypothesized radio technosignatures, take many different forms in terms of both the frequency width of the signal and the cadence of the signal. Each signal form requires a different search approach, and multiple methodologies have been explored over the years. Many of these approaches and signal forms were discussed in \cite{SETI2020}, but other techniques have been discussed since that time (i.e., \citealt{Houston_2021,Luan_2023,Suresh_2023}). However, publications on SETI have thus far placed a strong focus on drifting narrow frequency band (Hz-wide) signals (e.g., \citealt{enriquez2017turbo,Price_2020, Sheikh_2021, Ma_2023}).

COSMIC, at least initially, is designed to passively look for electromagnetic radiation at radio frequencies of 0.75--50\,GHz, which are observable by the standard VLA signal chain\footnote{The VLA has observation capabilities at frequencies as low as 74\,MHz, but these dipole receivers are rarely used and COSMIC is not currently operating using the frequencies between 0.074 and 0.75\,GHz. However, the data streams are available for use and COSMIC will be set up in the future to utilize them.}. To differentiate from the natural background of astrophysical emission, we search for continuous-wave (monochromatic) frequency-drifting signals in 1--8\,Hz-wide channels, where the drift comes from the difference in acceleration between the emitter and the receiver (the VLA telescope for COSMIC) \citep{2022ApJ...938....1L}.

\begin{figure*}
    \centering
    \includegraphics[width=\textwidth]{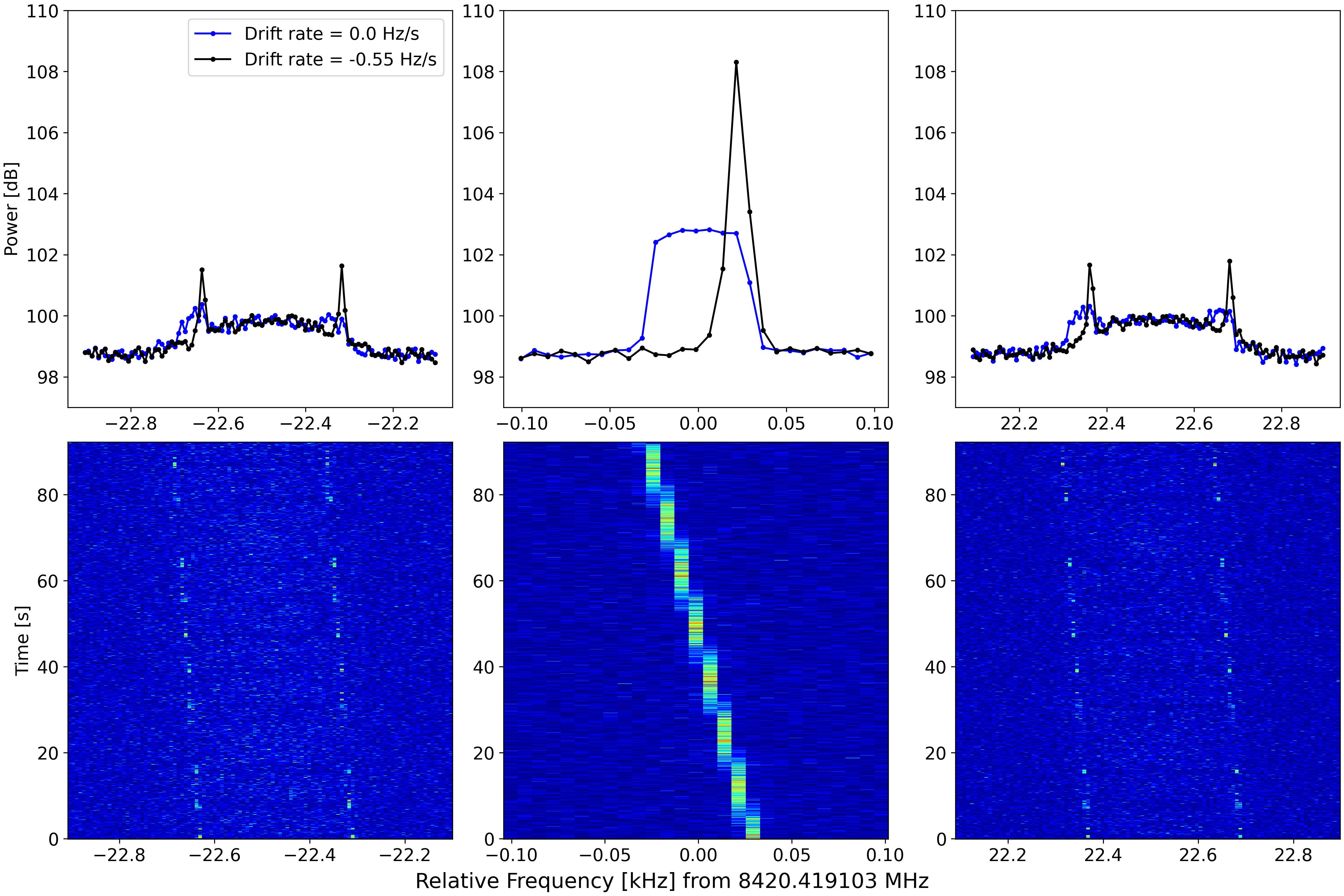}
\caption{The Voyager signal detected in targeted observations conducted with COSMIC. The upchannelization and beamforming were conducted at the same resolutions in frequency and time as those of the VLASS search mode (0.2\,s temporal resolution and 8\,Hz frequency resolution). The top row shows the integrated spectra without Doppler correction (green) and with Doppler correction (black). The bottom row shows the corresponding waterfall plots. The left and right columns show the sideband signals, and the middle column shows the carrier signal. The frequencies are recorded in the topocentric reference frame. }
\vspace{-0.5em}
    \label{Voyager}
\end{figure*}

As an example of a narrow-band technological signal and COSMIC's ability to detect such signals, we observed the Voyager 1 downlink signal sent to Earth at 8.4\,GHz and received by the VLA (Figure \ref{Voyager}). Voyager 1, launched on September 5, 1977, is currently 159\,AU from Earth within the constellation of Ophiuchus (within the Oort Cloud). The coordinates were obtained from the Horizons $astroquery$ database, and the expected frequency and Doppler shift were calculated using the \textsc{Python} $astropy$ package. COSMIC detected the signal in both the incoherent and coherent beams near the expected frequency. The coordinates to form the coherent beam were obtained based on images of Voyager 1 by correlating the raw voltages. Figure \ref{Voyager} shows the signal that was detected when a coherent beam was formed at the corresponding location. The autonomous real-time search pipeline also detected the signal and recorded the information in the database (a process described in more detail in \S 3.9).

The pipeline running in COSMIC uses the Doppler acceleration search algorithm \textsc{seticore}\footnote{\url{https://github.com/lacker/seticore}}, which is run on the data created via coherent and incoherent beamforming by means of the \textsc{BLADE} software package. \textsc{seticore} is a GPU implementation of \textsc{TurboSETI} \citep{enriquez2017turbo}, which is a Taylor tree search algorithm \citep{2019ApJ...884...14S} used in many narrow-band signal searches over the past five years. The user-provided input parameters for \textsc{seticore} include an SNR threshold, which was set to 10 for the first six months of COSMIC's operation, and a Doppler drift rate, which is currently set to $\pm$50\,Hz\,s$^{-1}$. However, these parameters can be changed within the observational \textsc{YAML} file.

At the end of a search, a series of information regarding the signals found (hits) is logged in a SQL database, and raw voltage ``stamp'' files for each antenna are saved in the storage nodes under a directory name that is a combination of the NRAO scheduling block ID, the project ID, and the scan ID. These stamp files are segments of the raw antenna voltages around the brightest signals, containing at least 200\,Hz of frequency data and all time samples available in the recording. These voltages can later be plotted using {\sc seticore} and/or correlated and imaged near signals of interest to confirm the signal characteristics. 

In the future, machine learning approaches similar to the search described in \cite{Ma_2023} could be applied to COSMIC-generated data. However, currently, the data are manually reviewed through statistical analyses, especially as we learn more about these data products. These analyses will be discussed in a future paper.

\subsection{Data Verification and Storage}
The COSMIC pipeline is designed to identify signals of interest by detecting narrow-band emission (``hits''), and then various post-processing pipelines may be created to filter those hits. In particular, two consecutive filters can be applied, a spatial filter (to determine whether a signal appears in one coherent beam but not in the others) and a drift rate filter (to determine whether the drift rate is between $\pm$ 50 Hz\,s$^{-1}$ but not identically zero, indicating a source that is accelerating relative to the receiver). In fact, the maximum drift rate is flexible and can be changed at any time depending on scientific interest. However, the current value of $\pm$50\,Hz\,s$^{-1}$ is set larger than that for a typical de-Doppler search for technosignatures \citep{2019ApJ...884...14S} due to our increased computing capabilities, allowing us to widen the restrictions on where we assume a transmitter to be located and how fast it is assumed to be rotating while also covering the drift rates determined by \cite{2022ApJ...938....1L} for the planetary systems nearest to our solar system.

\begin{figure}
    \centering
    \includegraphics[width = 0.49\textwidth]{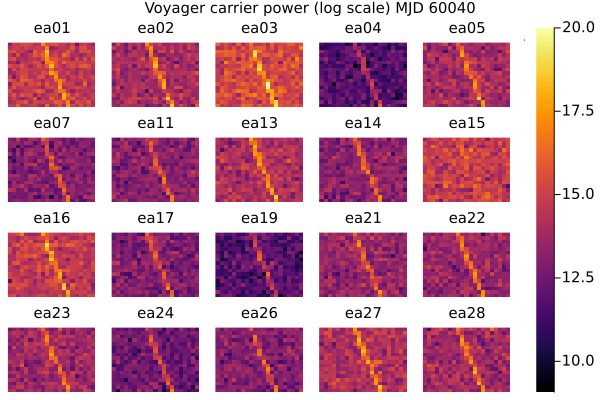}
\caption{The Voyager signal detected in the raw voltage ``stamp'' files for each online antenna. Antenna 15 (ea15) does not show a signal because it had a cryogenic problem and therefore was not producing data. The x-axis represents frequency, zoomed around the signal carrier frequency of 8420.573\,MHz (topocentric), and the y-axis represents time, with a total of 8\,s of data at a 0.2\,s resolution. The color scale represents the signal intensity in dB.}
\vspace{-0.5em}
    \label{Voyager_PS}
\end{figure}

Any real astronomical signal of interest would be a point source centered within a single coherent beam or with minimal leakage into closely packed nearby coherent beams. Terrestrial signals would, in contrast, ``bleed'' into multiple beams within a single FoV. The Doppler search kernel produces ``postage stamps'' of raw voltage files for at least 200\,Hz of bandwidth around the signal of interest and stores them in a series of folders for further evaluation, while the metadata regarding the hits contained within these stamp files are held in a SQL database. An example of the information that can potentially be obtained from such postage stamps is shown in Figure \ref{Voyager_PS}, where a drifting carrier signal is detected in the waterfall plots for each antenna. We would expect an astronomical signal to be detected by all operational antennas.

Given access to both coherent and incoherent beam data, we can assume that any signal present in a coherent beam will also appear in the incoherent beam, if the SNR is sufficiently high. Thus, we compare the incoherent beam with the coherent beam, the incoherent beam should contain a signal, as follows:

\begin{equation}
  coherent\_snr >  sqrt(N_{antennas})  * incoherent\_snr 
\end{equation}

If this is not true or if a signal is detected in all coherent beams, then the signal is likely to be RFI and should be ignored as an ETI signal of interest. However, as we investigate all information being recorded by COSMIC and processed through \textsc{BLADE} and \textsc{seticore}, all hits are currently recorded into a database without filtering. Instead, signal rejection based on this type of comparison between the coherent and incoherent beams is performed is a post-processing step, and the details will be provided in future work.

For any signal that passes these criteria, a check for known RFI or astronomical sources is completed, and a visual inspection of the characteristics of the drift profile is performed by means of waterfall plots. A total of ten steps are adopted by $Breakthrough$ $Listen$ to verify a signal, as explained in \cite{Sheikh_2021}, including checking the telescope and digital signal processing system for potential problems, examining the drift rate evolution and evaluation of the signal of interest, and searching for other instances of the signal of interest in archival data. If a signal passes all of these criteria, reobservation of the region from which the signal originated, with either the VLA or other telescopes, is critical as a check for redetection. If such a signal were to be confirmed, this would be revolutionary, allowing the field to blossom into a new era of understanding the possible opportunities for life outside our own planet.

Accordingly, we have successful target of opportunity (TOO) proposals in place with the VLA, the NRAO Very Long Baseline Array (VLBA), the Allen Telescope Array, and the Parkes 64 m telescope in Australia. Additionally, through collaboration with $Breakthrough$ $Listen$, we can utilize available time on the GBT. These collaborative efforts allow us to cover most, if not all, available frequencies that can be recorded with COSMIC and to cover the sky while utilizing telescopes with independent RFI environments and data collection systems. However, we make no assumption that a signal detected from an astronomical origin will be repeated. A repeated signal detected by another facility and/or hardware setup would be an ideal result. 

\subsection{System Monitoring}
COSMIC, as a system, spans many compute nodes, devices, and services and is reliant on all of them working in sync with the pipeline. Throughout the development of the software system, it was imperative for each process to log its state, errors, and interactivity, and this will continue to be important as additional functionalities are incorporated into COSMIC. These process logs allow faster debugging and shorten the time during which the system is not processing data. Where software services fail due to firmware or hardware issues, it becomes apparent that monitoring and logging the hardware and firmware status is also important.

The results of COSMIC delay (\S3.3.1), F-Engine (\S2.2), and calibration (\S3.3) state monitoring (\S3.4) are stored in an \textsc{Influx database}\footnote{\url{https://www.influxdata.com/}}, which supports the nanosecond-scale logging necessary when debugging delay tracking and F-Engine problems. Data are retained in this error database for only 30 days since the high read/write polling rates require it to be situated on the head node main drive, which has limited memory.

The COSMIC GPU compute node information is polled via \textsc{Prometheus}\footnote{\url{https://prometheus.io/}} exporters but is not stored for longer than a few hours due to storage resource limitations. \textsc{systemd}\footnote{\url{https://systemd.io/}} services are responsible for antenna control monitoring, while the delay engine, calibration, and \textsc{hashpipe} are also polled by different \textsc{Prometheus} exporters, as failure of these services will cause the system to stop commensal observations. It is not necessary to retain the compute node history, as this polling is primarily intended for alerting of compute problems requiring immediate attention.

Both the \textsc{Influx} content and \textsc{Prometheus} content are displayed via a \textsc{Grafana}\footnote{\url{https://grafana.com/}} dashboard. From this dashboard, an overview of the antenna, F-Engine, delay, calibration, and compute node processing states may be accessed. In addition, \textsc{Grafana} allows custom alerts to be configured to trigger on specific thresholds for any state change, and messages can be sent to a designated \textsc{Slack} channel.

\section{Commensal Observations with On-the-Fly Mapping}
In the on-the-fly mapping mode of VLASS, a slightly altered recording and target selection process is implemented in COSMIC. During VLASS observations, the telescope observes a continuous track in RA for approximately 10 minutes at a rate of 3.3\,arcminutes per second \citep{VLASS} before changing Dec and slewing in RA again. There is some overlap between these tracks, and each source is expected to be contained within the primary beam for approximately 5--8\,seconds.

\begin{figure}
    \centering
    \includegraphics[width = 0.45\textwidth]{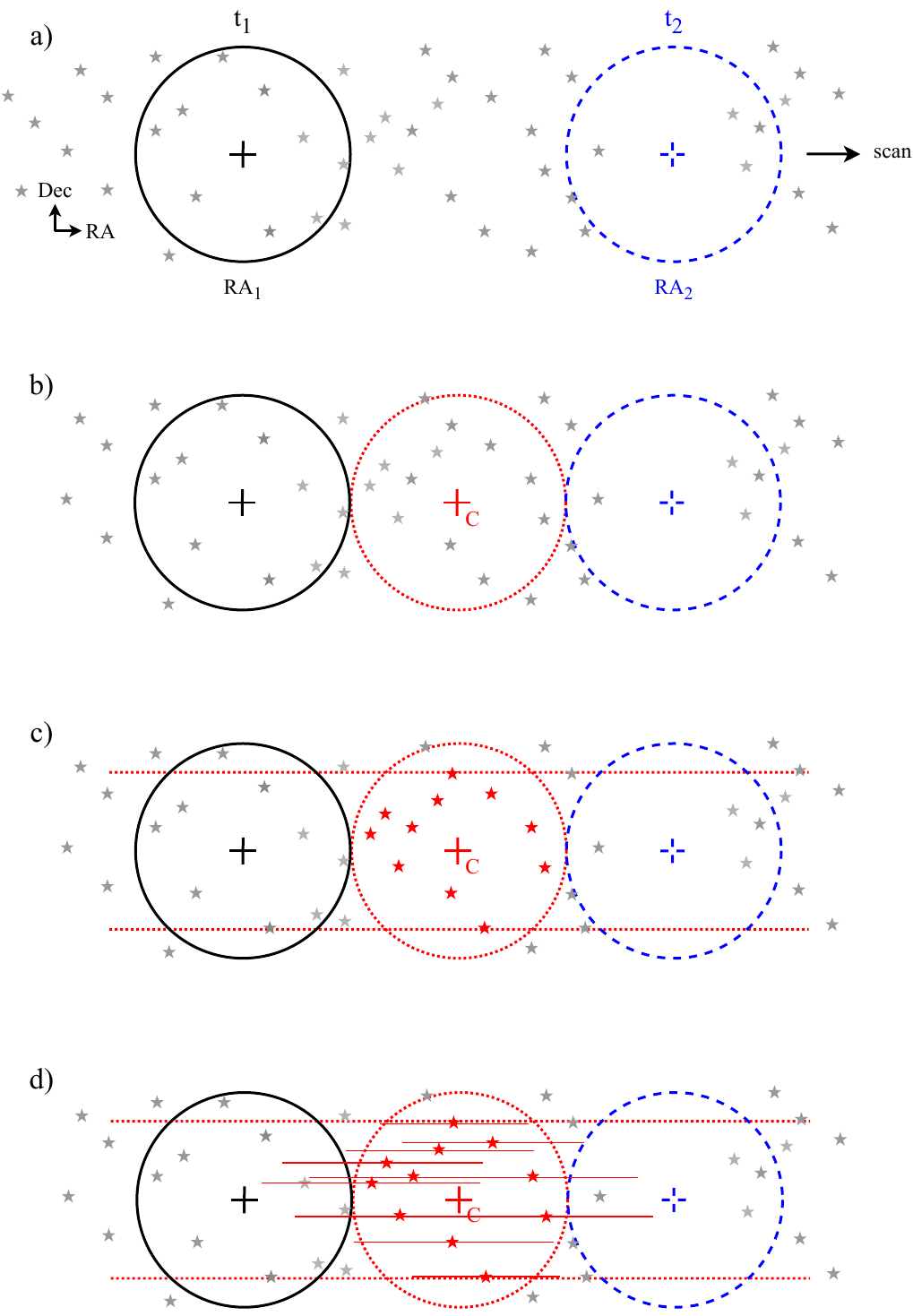}
\caption{A diagram illustrating the recording and target selection process during commensal observation with VLASS or other observations in on-the-fly mode. a) shows two circles representing the FoVs at the start of a recorded observation segment (black solid line) and at the end of the segment (green dashed line). These circles are spaced one FOV width apart, as illustrated in b), which shows the extrapolated FOV and phase center (\textit{C}, in red) for the recorded segment. c) shows the stars (solid red) that the target selector will choose as targets for coherent beamforming. d) approximately illustrates the track along which the chosen stars will appear to ``move'' within the primary FOV as it passes over them.}
\vspace{-0.5em}
    \label{fig:vlass_obs}
\end{figure}

To accommodate this, a simple design is implemented to handle the necessary data structures, with scope for increased complexity in the future. Figure \ref{fig:vlass_obs} shows the general outline of how the system handles data recording and processing. Each of the compute nodes receives a selection of 32 $\times$ 1\,MHz coarse channels containing approximately eight seconds of recorded data from each antenna.

In this case, the \textsc{hashpipe} automation directs the pipeline to record from \textit{time 1} to \textit{time 2} covering some FoV, as shown in part \textbf{a)} of Figure \ref{fig:vlass_obs}. As shown in \textbf{b)}, it also instructs the phase center to be located at \textit{C}, a central position between \textit{RA1} and \textit{RA2}, representing the RA coordinates of the phase center for the start and end of the recording segment. The target selector intelligently determines sources that should be in the field for at least five seconds and uses those as coherent beam targets, as illustrated in \textbf{c)} and \textbf{d)}. This process is repeated for each recorded time segment.

Not all stars within the circle around \textit{C} will be visible for the full width of the primary FoV. The horizontal red lines indicate the portion of the segment within which they will be recorded as the primary FoV passes over them. Stars further in declination from \textit{C} will be visible for shorter durations. We choose targets that will be observed for at least 4 seconds. The pipeline automation (via the target selector) writes the coordinates for the chosen sources (red stars in the diagram) into the $BFR5$ beamformer file such that the this file will contain the start and stop coordinates for each source. The beamformer is then instructed to coherently beamform toward each of the sources, either zeroing the coefficients when a source leaves the FoV or ignoring the corresponding data during this time.

In this current design, the pipeline automation deals with discrete recordings, each covering a transit equivalent to two primary beam widths. Each of these discrete recordings is completely handled in a separate $BFR5$ file. After recording and processing (fine channelization, beamforming, technosignature search, and cleanup), some sections of the sky may be missed, depending on the processing duration.

The primary goal of COSMIC processing during VLASS observations is to conduct narrow-band spectroscopy and Doppler acceleration searching on 5 coherent beams (with additional beams planned in the future, as discussed in \S 5.1) and an incoherent beam, providing a rate of $\sim$ 2000 sources per hour, with some sources observed multiple times. The maximum observation time for a given source can be assumed to be fixed at five seconds, which is governed by the VLASS observation strategy of constantly slewing the telescope to map a region of the sky. 

\section{Discussion}
One of the current goals of SETI is to gain an understanding of the prevalence of technologically advanced beings in the Universe, in particular through searches with radio telescopes. Because the electromagnetic emissions of most natural processes follow models of black-body radiation or synchrotron radiation with broad frequency features \citep{Cohen_1987}, it is expected that signals (at m, cm, or mm wavelengths) with a frequency resolution of less than 100\,Hz and down to the subhertz level are more likely to be a result of artificial radio generation. Although this may not be the case in all circumstances, as even on Earth it is not always true, looking for narrow-band signals is a probative technique that avoids some additional complications. In general, SETI is focused on discovering signals that do not have a readily available explanation for their existence. 

\begin{figure*}
    \centering
    \includegraphics[width = 0.85\textwidth]{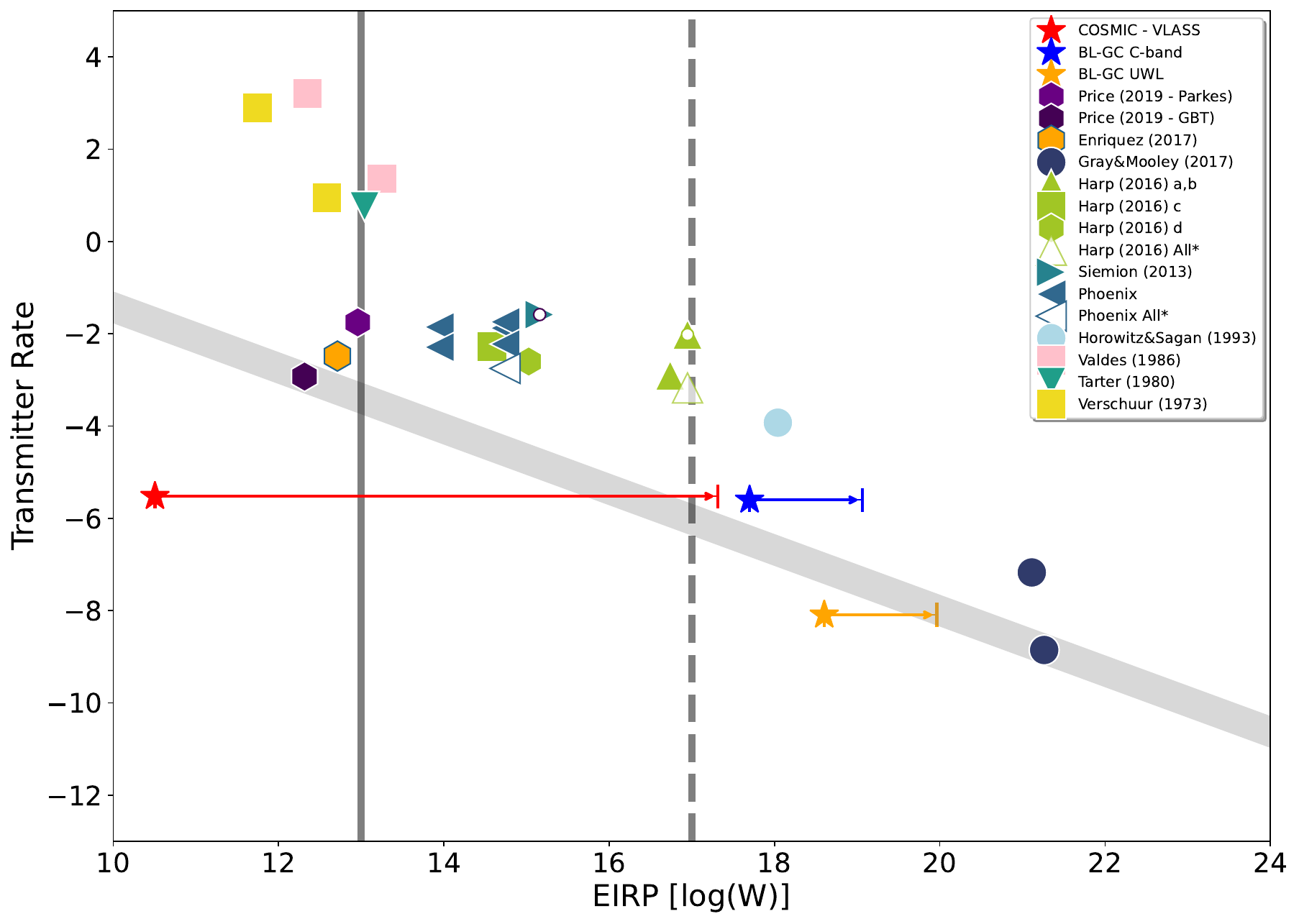}
\caption{A plot of previous SETI surveys covering frequencies greater than 1\,GHz. The x-axis represents the minimum detected equivalent isotropic radio power (EIRP;\citealt{Siemion_2013}). The y-axis represents the transmitter rate \citep{Price_2020}, which is indicative of the number of objects a survey observes. The COSMIC--VLASS point is based on the observations of previous epochs of the survey and the catalog of sources being used by the target selector. }
\vspace{-0.5em}
    \label{fig:EIRP}
\end{figure*}

Our main goal in COSMIC is to conduct an experiment to search for technosignatures that not only covers more sky area and frequencies than previous attempts but also is conducted in an effective and efficient manner. With COSMIC and its commensal abilities to observe during epoch 3 of VLASS during 2023 and 2024, we expect to target millions of stars at 2--4\,GHz and on the order of tens of thousands of stars during standard PI-driven science programs ranging in frequency from 0.75--50\,GHz in the same time frame. Therefore, the expectation is that COSMIC will search toward millions of stars at a sensitivity level at least comparable to that of Project Phoenix and be able to put well-constrained limits on the number of technologically advanced civilizations within our Galaxy, as shown in Figure \ref{fig:EIRP}. Currently, we are observing stars with coherent beams at a rate of 2000 stars per hour during VLASS. The exact rate of source observation during the full commensal mode along with PI-driven science has yet to be determined. This will be explained in more detail in a future paper focusing on the science output of COSMIC.

Thus far, however, we have already collected data in our database for over 400,000 sources from coherent beams, proving COSMIC to be a valuable and powerful machine for SETI.

\subsection{Future Plans for COSMIC}
The COSMIC system is flexibly designed to allow for future upgrades to hardware and software. By adding additional compute clusters, we can coherently beamform toward more targets, thus covering a larger number of stars or other astronomical sources of interest, or increasing the total bandwidth searched. With our current computing capabilities, we are searching five coherent beams when observing along with VLASS and between four and sixty-four beams for other observations, plus one incoherent beam. We can also record and search up to 1\,GHz of total bandwidth during VLASS and up to 1.4\,GHz of bandwidth along with other observations, although at the cost of reducing the number of coherent beams. In the future, we could increase the number of beams and the total bandwidth recorded and searched by expanding the computing infrastructure.

Currently, COSMIC is designed to work with the VLA's 8-bit system with a maximum total simultaneous bandwidth of 2\,GHz \citep{evla_11}. Future software upgrades will allow data to be recorded from the VLA when the 3-bit system is turned on, which offers 8\,GHz of simultaneous bandwidth. As mentioned in \S2, the channelization setup through the VLA control system does not impact COSMIC, so we can maintain flexibility regarding the resolutions in frequency and time with which we generate and record data. This flexibility will allow us to broaden the scientific output of COSMIC beyond SETI.

The current system is also designed to search for signals using the Stokes I total intensity. However, COSMIC ingests all four polarization data products and thus, with a software upgrade, could search for linear and circular polarization signals.

As with other programs where SETI is the main motivation for the design concept, other scientific endeavors can also be pursued with either a copy of the data or a change to the processing setup. In particular, COSMIC can enable searches for transients with submillisecond temporal resolution, such as fast radio bursts (e.g., \citealt{Faber_2021,Diermyer_2021}), or the data can enable spectral line science and axionic dark matter searches (e.g., \citealt{Foster_2022}). Moreover, with COSMIC's flexible design, many other scientific capabilities may be explored.

Another benefit of designing COSMIC as an Ethernet-based system is the ability to set up multicasting Ethernet technology for multimode commensal recording and processing systems. This means that other commensal systems could tap into the COSMIC digital processing rack to create other real-time scientific data outputs. Such initiatives could include offering additional functionalities for the VLA Low Band Ionospheric and Transient Experiment \citep{Clarke_VLITE}, $realfast$ \citep{Realfast,Law_2015}, or other commensal systems currently operating on the VLA.

\section{Conclusion}
COSMIC is a new digital backend on the VLA, built and designed to observe most of the Northern Hemisphere sky to look for signs of complex intelligent life through technosignatures. The computing system is designed using off-the-shelf digital components and is operated through open-source software. In the first development phase, COSMIC has been implemented with a real-time data processing pipeline to record digitized signals from the VLA, calibrate, perform coherent and incoherent beamforming, and search for narrow-band (1--8\,Hz) signals that may be consistent with non-natural signals of astronomical origin.

The commensal nature of COSMIC, paired with its autonomous real-time pipeline, allows it to achieve a remarkable leap forward in the search for extraterrestrial signals, beyond the constraints of previous programs dedicated to this purpose. Moreover, future developments of COSMIC could provide incredibly rich resources for the entire astronomical community.

\section{Software}
\begin{itemize}
\item {\sc topcat} -- \cite{Topcat}
\item {\sc CASA} -- \cite{CASA}
\item{NumPy v1.11.3 \citep{NumPy}, AstroPy \citep{Astropy}, SciPy \citep{SciPy}, Matplotlib \citep{Matplotlib}}, Pandas (\url{https://pandas.pydata.org/docs/})
\item {\sc pyUVdata} -- \cite{pyUVdata}
\item {\sc CARTA} -- \cite{angus_comrie_2020_3746095}
\item {\sc Blimpy} -- \url{https://github.com/zoips/blimpy}
\item {\sc ansible playbook} -- \url{https://docs.ansible.com/}
\item {\sc BLADE} --\url{https://github.com/luigifcruz/blade}
\item {\sc seticore}  -- \url{https://github.com/lacker/seticore}
\item {\sc hashpipe} -- \cite{MacMahon_2018}
\item xGPU -- \url{https://github.com/GPU-correlators/xGPU}
\item COSMIC Software -- \url{https://github.com/COSMIC-SETI}
\item target selector -- \url{https://github.com/danielczech/targets-minimal}
\item {\sc Redis} --\url{https://redis.com/}
\end{itemize}

\begin{acknowledgments}
In Memoriam: We thankfully acknowledge the work and assistance of Cindy George of the Servo-Fiber group at the VLA and Bart Wlodarczyk-Sroka, a PhD student at the University of Manchester, both of whom left us early in life and did not see the launch of science observations with COSMIC.

We gratefully acknowledge the foundational support from John and Carol Giannandrea that has made COSMIC possible. We acknowledge additional support from other donors including the Breakthrough Prize Foundation under the auspices of Breakthrough Listen. The National Radio Astronomy Observatory is a facility of the National Science Foundation operated under a cooperative agreement with Associated Universities, Inc.
\end{acknowledgments}

\vspace{5mm}
\facilities{VLA}

\bibliography{cosmic}{}
\bibliographystyle{aasjournal}

\end{document}